\documentclass[aps,prc,groupedaddress,twocolumn,nofootinbib]{revtex4-2}
\usepackage{bm}
\usepackage[dvipdfmx]{graphicx}
\usepackage{ascmac}
\usepackage{amsmath,amssymb}
\usepackage{cancel}
\usepackage{braket}
\usepackage{url}
\usepackage[setpagesize=false]{hyperref}
\usepackage{docmute}
\usepackage{calligra}
\usepackage{color}
\newcommand{\Jost}{\text{\textcalligra{f}\;}}
\allowdisplaybreaks[1]

\usepackage[normalem]{ulem}  

\begin{document}


\title{General amplitude of near-threshold hadron scattering for exotic hadrons}


\author{Katsuyoshi Sone}
\email[]{sone-katsuyoshi@ed.tmu.ac.jp}
\affiliation{Department of Physics, Tokyo Metropolitan University, Hachioji 192-0397, Japan}
\author{Tetsuo Hyodo}
\email[]{hyodo@tmu.ac.jp}
\affiliation{Department of Physics, Tokyo Metropolitan University, Hachioji 192-0397, Japan}


\date{\today}

\begin{abstract}
We discuss the general behavior of the scattering amplitude with channel couplings near the two-body threshold. It is known that the Flatt\'{e} amplitude, which is often used in the analysis of experimental data involving exotic hadrons, has some constraint in the near-threshold energy region. While the M-matrix gives the general expression of the scattering amplitude, it is not smoothly connected to the Flatt\'{e} amplitude, due to the property of the determinant of the amplitude in channel space. In this paper, based on the effective field theory, we propose new parametrization of the scattering amplitude which gives the general expression near the threshold and has a well-defined limit reproducing the Flatt\'{e} amplitude. We show that the nonresonant background contribution exists in the general amplitude even in the first order in the momentum. Finally, we quantitatively evaluate the cross sections by changing the strength of the background contribution. 
We find that the interference with the background term may induce a dip structure of the cross section near the threshold, in addition to the peak and threshold cusp structures.
\end{abstract}


\maketitle

\section{Introduction}
Exotic hadrons are composed of different combinations of quarks from mesons and baryons, and their internal structure still remains unresolved. In recent years, experimental progress has led to the discovery of a wide variety of the exotic hadrons, and various studies have been conducted from both the theoretical and experimental perspectives~\cite{Guo:2017jvc, Brambilla:2019esw}. Most exotic hadrons are unstable states with finite lifetimes and decay into multiple hadrons, and therefore they are observed as resonance peaks in the hadron scattering. Accumulation of the experimental data reveals that a lot of exotic hadrons appear near the threshold. Furthermore, in actual hadron systems, the channel couplings play an important role by inducing the inelastic scattering in addition to the elastic scattering. As an example, $f_0(980)$ appears near the $K\bar{K}$ threshold and decays into the $\pi\pi$ channel, so we need to consider the coupled-channel scattering of the $\pi\pi$-$K\bar{K}$ system~\cite{ParticleDataGroup:2022pth}.

In the scattering amplitude, the eigenenergy of the resonance state is represented as a pole in the complex energy plane~\cite{Taylor,Hyodo:2020czb}. To express the scattering amplitude with a resonance pole, one utilizes the Breit-Wigner amplitude having a pole at the energy $E=E_{\rm R}-i\Gamma/2$, where $E_{\rm R}$ represents the resonance energy and $\Gamma$ the decay width. Because the Breit-Wigner amplitude with a constant decay width $\Gamma$ does not include the effect of the threshold, the Breit-Wigner amplitude cannot be applied to the analysis of the resonance state near the threshold. To incorporate the threshold effect into the Breit-Wigner amplitude, the Flatt\'{e} amplitude~\cite{Flatte:1976xu} has been frequently used for the actual analysis of the various hadron scattering near the threshold~\cite{Bugg:1994mg,E852:1996san,Achasov:2000ku,Achasov:2000ym,FOCUS:2004prc,BES:2004twe,KLOE:2005jxf,Belle:2005rpz,KLOE:2006vmv,Bugg:2008ig,KLOE:2009ehb,CDF:2011kjt,CLEO:2011upl,LHCb:2020xds}. In the Flatt\'{e} amplitude, the opening of the threshold is taken into account by the energy dependence of the decay width $\Gamma$.

To study the near-threshold exotic hadrons, it is useful to expand the inverse of the scattering amplitude as the power series of the momentum $k$ of the threshold channel. In the single-channel case, the effective range expansion of the scattering amplitude $f(k)$ reads
\begin{align}
    f(k)=\frac{1}{-\frac{1}{a} + \frac{r}{2}k^2 + {\cal O}(k^4) - ik}, \label{eq: 1 ERE}
\end{align}
where $a$ and $r$ are called the scattering length and effective range, respectively. It is important to determine $a$ and $r$ because the near-threshold dynamics, including the position of the near-threshold pole, is highly constrained by these constants. In the coupled-channel scattering, it is known that the denominator of the Flatt\'{e} amplitude can also be written in the form of the effective range expansion~\cite{Baru:2004xg}. This feature allows one to use the Flatt\'{e} amplitude to determine the scattering length and effective range~\cite{Esposito:2021vhu, LHCb:2021auc} (see also the discussion in Ref.~\cite{Baru:2021ldu}). 

However, a problem of the Flatt\'{e} amplitude has been pointed out; the number of the independent parameters decreases near the threshold~\cite{Baru:2004xg}. This fact suggests that some conditions are imposed on the Flatt\'{e} amplitude near the threshold, lacking the generality of its expression. On the other hand, the general expression of the scattering amplitude consistent with the optical theorem has been constructed, for instance, by using the M-matrix approach~\cite{Badalian:1981xj}. As we will show below, however, the M-matrix type amplitude cannot be directly reduced to the Flatt\'{e} amplitude due to the implicit assumption in its derivation. To study how the general expression gradually approaches the Flatt\'{e} amplitude, it is desirable to construct an alternative amplitude, which keeps the generality of the expression but having the smooth connection with the Flatt\'{e} amplitude.

In this study, we focus on the above-mentioned problems of the near-threshold coupled-channel scattering amplitude and clarify the relation between the general form of the amplitude and the Flatt\'{e} amplitude. For this purpose, we formulate both the Flatt\'{e} and M-matrix amplitudes in the framework of the effective field theory. Through the detailed comparison of these amplitudes, we propose a new parametrization of the scattering amplitude that does not lose generality even near the threshold and directly reduces to the Flatt\'{e} amplitude. This allows us to study how the general behaviors of the scattering length and near-threshold cross sections reduce to those of the Flatt\'{e} amplitude.

This paper is organized as follows. First, we derive the  M-matrix type amplitude (hereafter called the Contact amplitude) and the Flatt\'{e} amplitude from the effective field theory in Sec.~\ref{sec : coupled channel}. We then compare the Flatt\'{e} amplitude with the Contact amplitude and clarify the imposed conditions by exploring the compatibility between these amplitudes. Next, in Sec.~\ref{sec : General amplitude}, we propose a new parametrization of the scattering amplitude (General amplitude) that unifies the Contact amplitude and the Flatt\'{e} amplitude. Using the General amplitude, we discuss in detail the nature of the scattering amplitude near the threshold. In Sec.~\ref{sec: cs fg} the behavior of the scattering cross section near the threshold is quantitatively investigated using the General amplitude. A summary is given in the last section. Preliminary results of Sec.~\ref{sec : General amplitude} are partly reported in the proceedings of the conference~\cite{Sone:2023fol}.

\section{Coupled-channel scattering amplitude}\label{sec : coupled channel}

In this section, based on Refs.~\cite{Cohen:2004kf, Braaten:2007nq,Dong:2020hxe}, we review the derivation of the Contact amplitude and Flatt\'{e} amplitude for two-channel scattering from the effective field theory (EFT) in Sec.~\ref{subsec: derivation of Contact} and in Sec.~\ref{subsec : Flatte amplitude}, respectively. We discuss the scattering length and effective range in these amplitudes. In Sec.~\ref{subsec: comparison C and F}, we show that the Contact amplitude does not directly reduce to the Flatt\'{e} amplitude focusing on the determinant of the scattering amplitude.

\subsection{Contact amplitude}\label{subsec: derivation of Contact}
Here we derive the two-channel scattering amplitudes at low energies for systems with contact four-point interaction from effective field theory. The Lagrangian $\mathcal{L}$ in this case is given by~\cite{Cohen:2004kf, Braaten:2007nq,Dong:2020hxe}
\begin{align}
    \mathcal{L}_{\rm C} &= \mathcal{L}_{0}^{\rm C} + \mathcal{L}_{\rm int}^{\rm C} , \\
    \mathcal{L}_{0}^{\rm C} &= \psi^{\dagger}_{1}\left(i \partial_0 + \frac{\nabla^2}{2m_1} - (m_1 - m_2)\right)\psi_1 \notag \\
    &\quad + \phi_1^{\dagger}\left(i \partial_0 + \frac{\nabla^2}{2M_1} - (M_1 - M_2)\right)\phi_1 \notag \\
    &\quad + \psi_2^{\dagger}\left(i \partial_0 + \frac{\nabla^2}{2m_2}\right)\psi_2 + \phi_2^{\dagger}\left(i \partial_0 + \frac{\nabla^2}{2M_2}\right)\phi_2, \\
    \mathcal{L}_{\rm int}^{\rm C} &= -c_{ij}\psi^{\dagger}_{i}\phi^{\dagger}_{i}\phi_j\psi_j + \cdots, \label{eq: Lint}
\end{align}
where $\mathcal{L}_{0}^{\rm C}$ is the free Lagrangian representing the rest energy and non-relativistic kinetic energy of the particles. $\psi_i$ and $\phi_i\ (i=1,2)$ are the fields corresponding to the two particles in channel $i$ and $m_i$ and $M_i\ (i=1,2)$ represent the masses of the particles in channel $i$. Since the energy is measured from the threshold of channel 2 in this study, the free Lagrangian $\mathcal{L}_{0}^{\rm C}$ includes the contribution of the rest energies of $\psi_1$ and $\phi_1$. $\mathcal{L}_{\rm int}^{\rm C}$ is the Lagrangian of the four-point interaction that represents the transition between the  $\psi_{i}\phi_i$ and $\psi_{j}\phi_j$ scattering channels, and $c_{ij}$ is the coupling constant. The higher order terms in the derivative expansion in $\mathcal{L}_{\rm int}^{\rm C}$ are abbreviated. We consider the low-energy scattering, where the contact interaction in Eq.~\eqref{eq: Lint} dominates and the higher-order terms are assumed to be negligible.

The Feynman rules obtained from the Lagrangian $\mathcal{L}_{\rm C}$ give the vertex $V_{\rm C}$ as
\begin{align}
    V_{\rm C} = 
    \begin{pmatrix}
        c_{11} & c_{12} \\
        c_{12} & c_{22}
    \end{pmatrix},
    \label{eq:EFT effective potential}
\end{align}
with the bare parameters $c_{11},c_{12}$, and $c_{22}$. The $(i,j)$ components of $V_{\rm C}$ describe the direct transition process from the scattering channel $j$ to  $i$. Since we are interested in the two-body scattering, we consider the four-point function $T_{ij}$ by adding up all possible diagrams for the two-body to two-body process obtained from the Feynman rules. In the present case, the four-point function is given only by the diagrams obtained from the Lippmann-Schwinger equation, in which $V_{\rm C}$ is regarded as a potential. The resulting four-point function, the T-matrix of this scattering, depends only on the total energy $E$ in the center-of-mass system:
\begin{align}
    [T(E)]^{-1} &= [V_{\rm C}]^{-1} - G^{\varLambda}(E), \label{eq:the matrix Lippmann-Schwinger V depends on E}
\end{align}
where
\begin{align}
    G^{\varLambda}(E) &=
    \begin{pmatrix}
        G^{\varLambda}_1(E) & 0 \\
        0 & G^{\varLambda}_{2}(E)
    \end{pmatrix}\notag \\
    &= 
    \begin{pmatrix}
        - \frac{\mu_1\varLambda}{\pi^2} - \frac{\mu_1}{2\pi}ip & 0\\
        0 & - \frac{\mu_2\varLambda}{\pi^2} - \frac{\mu_2}{2\pi}ik
    \end{pmatrix}, \label{eq: G varl} \\
    G^{\varLambda}_i(E)
    &=\int\frac{d\bm{q}}{(2\pi)^3}
    \frac{1}{E+\Delta \delta_{i1}-\bm{q}^2/(2\mu_i)+i0^+},
\end{align}
where $\Delta=(M_2+m_2)-(M_1+m_1)$ is the threshold energy difference, $\mu_i=M_im_i/(M_i+m_i)$ the reduced mass in channel $i$, and $p=\sqrt{2\mu_1(E + \Delta)}, k=\sqrt{2\mu_2E}$ the relative momenta of channels 1 and 2, respectively. Since the potential $V_{\rm C}$ is a contact interaction, we have introduced the cutoff $\varLambda$, the upper limit of the momentum integral to regularize $G^{\varLambda}(E)$, by keeping the leading contribution in the $\varLambda\to \infty$ limit in Eq.~\eqref{eq: G varl}.

Substituting $V_{\rm C}$ into Eq.~\eqref{eq:the matrix Lippmann-Schwinger V depends on E}, we obtain 
\begin{align}
    &\quad [T(E)]^{-1} \notag \\
    &=
    \begin{pmatrix}
            \frac{c_{22}}{c_{11}c_{22}-c_{12}^2} + \frac{\mu_1 \varLambda}{\pi^2} + \frac{\mu_1}{2\pi}ip & -\frac{c_{12}}{c_{11}c_{22}-c_{12}^2} \\
            -\frac{c_{12}}{c_{11}c_{22}-c_{12}^2} & \frac{c_{11}}{c_{11}c_{22}-c_{12}^2} + \frac{\mu_2 \varLambda}{\pi^2} + \frac{\mu_2}{2\pi}ik
    \end{pmatrix}.\label{eq:the T-matrix represented by lambda}
\end{align}
To perform the renormalization, we define the physical quantities $a_{11},a_{12}$, and $a_{22}$ so that the T-matrix takes the form
\begin{align}
    [T(E)]^{-1} = 
    \begin{pmatrix}
        \frac{\mu_1}{2\pi}\frac{1}{a_{11}} + \frac{\mu_1}{2\pi}ip & -\frac{\sqrt{\mu_1\mu_2}}{2\pi}\frac{1}{a_{12}} \\
        -\frac{\sqrt{\mu_1\mu_2}}{2\pi}\frac{1}{a_{12}} & \frac{\mu_2}{2\pi}\frac{1}{a_{22}} + \frac{\mu_2}{2\pi}ik
    \end{pmatrix}.
    \label{eq:the invers matrix of the T-matrix}
\end{align}
Comparing Eq.~\eqref{eq:the invers matrix of the T-matrix} with Eq.~\eqref{eq:the T-matrix represented by lambda}, we introduce the cutoff $\varLambda$ dependence in the bare parameters $c_{11},c_{12}$, and $c_{22}$ so that $a_{11},a_{12}$, and $a_{22}$ are cuttoff-independent:
\begin{align}
    \frac{c_{22}(\varLambda)}{c_{11}(\varLambda)c_{22}(\varLambda) - c_{12}^2(\varLambda)} + \frac{\mu_1 \varLambda}{\pi^2} &= \frac{\mu_1}{2\pi}\frac{1}{a_{11}}, \label{eq:a_{11} kurikomi}\\
    \frac{c_{11}(\varLambda)}{c_{11}(\varLambda)c_{22}(\varLambda) - c_{12}^2(\varLambda)} + \frac{\mu_2 \varLambda}{\pi^2} &= \frac{\mu_2}{2\pi}\frac{1}{a_{22}}, \label{eq:a_{22} kurikomi}\\
    \frac{c_{12}(\varLambda)}{c_{11}(\varLambda)c_{22}(\varLambda) - c_{12}^2(\varLambda)} &= \frac{\sqrt{\mu_1 \mu_2}}{2\pi} \frac{1}{a_{12}}. \label{eq:a_{12} kurikomi}
\end{align}
Taking the $\varLambda \rightarrow \infty$ limit with the physical quantities $a_{11},a_{12}$, and $a_{22}$ kept finite, the inverse matrix of the scattering amplitude is given by 
\begin{align}
    [f^{\rm C}(E)]^{-1}
    = 
    \begin{pmatrix}
        -\frac{1}{a_{11}} - ip & \frac{1}{a_{12}} \\
        \frac{1}{a_{12}} & -\frac{1}{a_{22}} - ik
    \end{pmatrix},
    \label{eq:the invers matrix of the EFT amplitude}
\end{align}
where the relation between the T-matrix and scattering amplitude $f_{ij}=-\sqrt{\mu_i\mu_j}T_{ij}/(2\pi)$ is used. From Eq.~\eqref{eq:the invers matrix of the EFT amplitude}, we obtain the scattering amplitude
\begin{align}
    f^{\rm C}(E) &= \frac{1}{\frac{1}{a_{12}^2} - \left(\frac{1}{a_{11}} + ip\right)\left(\frac{1}{a_{22}} + ik\right)} \notag \\
    &\quad \times
    \begin{pmatrix}
        \frac{1}{a_{22}} + ik & \frac{1}{a_{12}} \\
        \frac{1}{a_{12}} & \frac{1}{a_{11}} + ip
    \end{pmatrix}.
    \label{eq:EFT amplitude}
\end{align}
Hereafter, we call this amplitude $f^{\rm C}(E)$ the Contact amplitude. 

It is instructive to discuss the relation between the Contact amplitude and the two-channel M-matrix type scattering amplitude derived from the optical theorem. The general form of the inverse of the two-channel scattering amplitude is given as
\begin{align}
    [f(E)]^{-1}=
    \begin{pmatrix}
        M_{11}(E) - ip & M_{12}(E) \\
        M_{12}(E) & M_{22}(E) - ik
    \end{pmatrix}, \label{eq: two channel scattering amplitude}
\end{align}
using the M-matrix $M_{ij}(E)$, which is a function of the energy~\cite{Badalian:1981xj}. In Eq.~\eqref{eq: two channel scattering amplitude}, by setting $M_{ij}(E)$ as a constant $M_{ij} = 1/a_{ij}$, we recover the Contact amplitude~\eqref{eq:the invers matrix of the EFT amplitude}. In other words, it can be seen that the Contact amplitude $f^{\rm C}(E)$ in Eq.~\eqref{eq:EFT amplitude} is a special case of the M-matrix type scattering amplitude. This ensures that $f^C(E)$ is a general low-energy scattering amplitude consistent with the optical theorem near the threshold where the higher-order terms in energy $E$ can be neglected. From Eq.~\eqref{eq:EFT amplitude}, we find that the general form of the two-channel scattering amplitude is characterized by three parameters $a_{11},a_{12}$, and $a_{22}$ near the threshold. In general, by imposing the time-reversal symmetry on the M-matrix representation, the off-diagonal components of the M-matrix becomes $M_{ij}=M_{ji}$, so the $N$ channel scattering amplitude has $N(N+1)/2$ independent parameters near the threshold (see Appendix~\ref{sec:Nchannel}).

We now discuss the scattering length and effective range at the threshold of channel 2 (higher energy threshold) using the Constant amplitude in Eq.~\eqref{eq:EFT amplitude}. Expanding the denominator of the (2,2) component of the scattering amplitude $f^{\rm C}_{22}(E)$ 
in terms of the momentum $k$, we obtain 
\begin{align}
    f^{\rm C}_{22}(E) 
    &= 
    \Biggl[\frac{1}{a_{12}^2\left(\frac{1}{a_{11}} + ip_0\right)} - \frac{1}{a_{22}}  \notag \\
    &\quad - \frac{i}{2a_{12}^2\left(\frac{1}{a_{11}} + ip_0\right)^2p_0}k^2 - ik +{\cal O}(k^4)\Biggr]^{-1}, \label{eq:ERE of 22 component of the EFT amplitude}
\end{align}
where $p_0=\sqrt{2\mu_1\Delta}$ represents the momentum of channel 1 at the threshold of channel 2. The denominator of $f^{\rm C}_{22}(E)$ is composed only of the even powers in $k$, except for the term $-ik$. Therefore, $f^{\rm C}_{22}(E)$ can be written in the form of the effective range expansion in $k$, the momentum of channel 2. As in the case of the single channel scattering~\eqref{eq: 1 ERE}, we define the scattering length $a_{\rm C}$ from the constant term and the effective range $r_{\rm C}$ from the coefficient of the $k^2$ term in the denominator of $f^{\rm C}_{22}(E)$ in Eq.~\eqref{eq:ERE of 22 component of the EFT amplitude}. The scattering length $a_{\rm C}$ is given by
\begin{align}
    a_{\rm C} &= \frac{a_{12}^2a_{22}\left(1 + ip_0a_{11}\right)}{a_{12}^2\left(1 + ip_0a_{11}\right) - a_{11}a_{22}}.
    \label{eq:the scattering length of the EFT amplitude}
\end{align}
For later discussion, we decompose $a_{\rm C}$ into the real and imaginary parts:
\begin{align}
    a_{\rm C} &= \frac{a_{12}^4a_{22}(1+p_0^2a_{11}^2) - a_{11}a_{12}^2a_{22}^2}{(a_{12}^2 - a_{11}a_{22})^2 + p_0^2 a_{11}^2a_{12}^4} \notag \\
    &\quad - i\frac{p_0a_{11}^2a_{12}^2a_{22}^2}{(a_{12}^2 - a_{11}a_{22})^2 + p_0^2 a_{11}^2a_{12}^4}. \label{eq: ReaEFT + iImaEFT}
\end{align}
This expression confirms the relation ${\rm Im}(a_{\rm C})<0$ required by the optical theorem. Similarly, the effective range $r_{\rm C}$ is obtained as
\begin{align}
    r_{\rm C} &=  -\frac{i}{p_0} \left\lbrace \frac{a_{11}}{a_{12} \left( 1 + ip_0a_{11} \right)} \right\rbrace^2.
    \label{eq:the effective range of the EFT amplitude}
\end{align}
Note that both $a_{\rm C}$ and $r_{\rm C}$ are in general complex, reflecting the effect of the decay into channel 1.
We also expand the denominator of the (1,1) component of the scattering amplitude $f^{\rm C}_{11}(E)$
(lower energy channel)
in terms of $k$
\begin{align}
    f^{\rm C}_{11}(E) &= \frac{a_{12}^2}{a_{22}^2}\left[\frac{1}{a_{22}} - \frac{a_{12}^2}{a_{11}a_{22}^2} - i\frac{a_{12}^2}{a_{22}^2}p_0 \right. \notag \\
    &\quad \left. - \left(a_{22} + i\frac{a_{12}^2}{2a_{22}^2p_0}\right) k^2 - ik + {\cal O}(k^3)\right]^{-1}. \label{eq:ERE of 11 component of the EFT amplitude}
\end{align}
In contrast to $f^{\rm C}_{22}(E)$, the denominator of $f^{\rm C}_{11}(E)$ contains the terms of $k^3$ or more odd powers. Therefore, $f^{\rm C}_{11}(E)$ cannot be written in the form of the effective range expansion in $k$, and the scattering length and effective range should be defined not in $f^{\rm C}_{11}(E)$ but in $f^{\rm C}_{22}(E)$.

We comment on the higher order corrections to the effective range of the Contact amplitude $r_{\rm C}$ in Eq.~\eqref{eq:the effective range of the EFT amplitude}, which is written only by the parameters $a_{11},a_{12}$, and $a_{22}$. Considering the higher order terms of the interaction Lagrangian in the derivative expansion, one can enlarge the applicable energy region of the EFT.
The inclusion of the next-to-leading order terms gives the scattering amplitude
\begin{align}
    [f^{\rm C}(E)]^{-1} \simeq 
    \begin{pmatrix}
        -\frac{1}{a_{11}} + a^{(2)}_{11}k^2 - ip & \frac{1}{a_{12}} + a_{12}^{(2)}k^2 \\
        \frac{1}{a_{12}} + a_{12}^{(2)}k^2 & -\frac{1}{a_{22}} + a_{22}^{(2)}k^2 - ik
    \end{pmatrix}, \label{eq: the EFT amplitude expansion parameter}
\end{align}
where $a_{11}^{(2)},a_{12}^{(2)}$ and $a_{22}^{(2)}$ are the coefficients of the $k^2$ terms. The scattering length  $a_{\rm C}$ determined by $f^{\rm C}_{22}(E)$ in Eq.~\eqref{eq: the EFT amplitude expansion parameter} remains unchanged from the expression in Eq.~\eqref{eq:the scattering length of the EFT amplitude}, i.e., $a_{\rm C}$ does not suffer from the higher order corrections. In contrast, the effective range $r_{\rm C}$ contains the higher order corrections with $a_{11}^{(2)},a_{12}^{(2)}$ and $a_{22}^{(2)}$. Thus, the expression of $r_{\rm C}$ in Eq.~\eqref{eq:the effective range of the EFT amplitude} is valid only in the leading order EFT, and the value can be modified by including the higher order corrections.

\subsection{Flatt\'{e} amplitude}\label{subsec : Flatte amplitude}

In this section, we first show the derivation of the Flatt\'{e} amplitude in the EFT. The Lagrangian that gives the Flatt\'{e} amplitude is~\cite{Kinugawa:2023fbf}
\begin{align}
    \mathcal{L}_{\rm F} &= \mathcal{L}^{\rm F}_{0} + \mathcal{L}^{\rm F}_{\rm int}, \\
    \mathcal{L}^{\rm F}_{0} &= \psi^{\dagger}_{1}\left(i \partial_0 + \frac{\nabla^2}{2m_1} - (m_1 - m_2)\right)\psi_1 \notag \\
    &\quad + \phi_1^{\dagger}\left(i \partial_0 + \frac{\nabla^2}{2M_1} - (M_1 - M_2)\right)\phi_1 \notag \\
    &\quad+ \psi_2^{\dagger}\left(i \partial_0 + \frac{\nabla^2}{2m_2}\right)\psi_2 + \phi_2^{\dagger}\left(i \partial_0 + \frac{\nabla^2}{2M_2}\right)\phi_2 \notag \\
    &\quad+ \Psi^{\dagger} \left(i\partial_0 + \frac{\nabla^2}{2M}-\nu\right)\Psi, \\
    \mathcal{L}^{\rm F}_{\rm int} &= \hat{g}_{i} \left(\psi^{\dagger}_i\phi^{\dagger}_i\Psi + \Psi^{\dagger}\phi_i\psi_i\right),
\end{align}
where $\mathcal{L}^{\rm F}_{0}$ is the free Lagrangian  and $\mathcal{L}^{\rm F}_{\rm int}$ is the interaction Lagrangian. In addition to the $\psi_i$ and $\phi_i$ fields, we introduce the bare field $\Psi$ with mass $M$. The parameter $\nu=M - (m_2+M_2)$ represents the bare energy measured from the threshold of channel 2. The parameter $\hat{g}_i$ stands for the coupling constant between the scattering state of channel $i$ and the bare field.

From the Lagrangian $\mathcal{L}_{\rm F}$, the coupled-channel potential $V_{\rm F}(E)$ for the two-body to two-body process is given by  
\begin{align}
    V_{\rm F}(E) &= \frac{1}{E - \nu}
    \begin{pmatrix}
        \hat{g}_{1}^2 & \hat{g}_1\hat{g}_2\\
        \hat{g}_1\hat{g}_2 & \hat{g}_{2}^2
    \end{pmatrix}, \label{eq: the effective potential for Flatte E}
\end{align}
using the bare parameters $\nu, \hat{g}_1$, and $\hat{g}_2$. The potential $V_{\rm F}(E)$ describes the transition between the scattering states through the s-channel exchange of the bare field $\Psi$. It follows from the expression~\eqref{eq: the effective potential for Flatte E} that $\det[V_{\rm F}(E)]=0$, and therefore $V_{\rm F}(E)$ does not have the inverse matrix. Hence, $V_{\rm F}$ cannot be directly substituted into the expression~\eqref{eq:the matrix Lippmann-Schwinger V depends on E} and we use an equivalent equation without $V^{-1}$:
\begin{align}
    T(E) &= [\hat{1} - V(E) G^{\varLambda}(E)]^{-1} V(E). \label{eq: Flatte you lippman}
\end{align}
Substituting $V_{\rm F}(E)$ into Eq.~\eqref{eq: Flatte you lippman}, we obtain the T-matrix
\begin{align}
    T(E) &= \left[E - \nu + \hat{g}_1^2 \left( \frac{\mu_1\varLambda}{\pi^2} + \frac{\mu_1}{2\pi}ip\right) \right. \notag \\
    &\quad \left.+ \hat{g}_2^2 \left(\frac{\mu_2\varLambda}{\pi^2} + \frac{\mu_2}{2\pi}ik\right) \right]^{-1}
    \begin{pmatrix}
        \hat{g}_1^2 & \hat{g}_1\hat{g}_2 \\
        \hat{g}_1\hat{g}_2 &  \hat{g}_2^2
    \end{pmatrix}. \label{eq: the T-matrix for Flatte before kurikomi}
\end{align}
To perform the renormalization, we define the physical quantities $E_{\rm BW},g_1$, and $g_2$ so that the T-matrix takes the form 
\begin{align}
    T(E) &= \frac{1}{2E - 2E_{\rm BW} + ig_1^2p + ig_2^2k} \notag \\
    &\quad \times
    \begin{pmatrix}
        \frac{2\pi}{\mu_1}g_1^2 & \frac{2\pi}{\sqrt{\mu_1\mu_2}}g_1g_2 \\
        \frac{2\pi}{\sqrt{\mu_1\mu_2}}g_1g_2 & \frac{2\pi}{\mu_2}g_2^2
    \end{pmatrix}. \label{eq: Tmat Flatte E}
\end{align}
Comparing Eq.~\eqref{eq: the T-matrix for Flatte before kurikomi} with Eq.~\eqref{eq: Tmat Flatte E}, we introduce the cutoff $\varLambda$ dependence in the bare parameters $\nu,\hat{g_1}$, and $\hat{g_2}$ so that $E_{\rm BW},g_1$, and $g_2$ are cuttoff-independent:
\begin{align}
    \nu(\varLambda) - g_1^2\frac{\varLambda}{\pi} - g_2^2\frac{\varLambda}{\pi} &= E_{\rm BW} , \label{eq: nu kurikomi} \\
    \hat{g}_{1}(\varLambda) &= \sqrt{\frac{\pi}{\mu_1}}g_1, \label{eq: g1 kurikomi} \\
    \hat{g}_2(\varLambda) &= \sqrt{\frac{\pi}{\mu_2}}g_2. \label{eq: kurikomi}
\end{align}
Note that there is no $\varLambda$-dependence in the coupling constants $\hat{g}_1$ and $\hat{g}_2$. Taking the $\varLambda \rightarrow \infty$ limit with the physical quantity $E_{\rm BW}$ kept finite, the Flatt\'{e} scattering amplitude is obtained as 
\begin{align}
    f^{\rm F}(E) &= \frac{1}{2E_{\rm BW} - 2E - ig_1^2p - ig_2^2k} \notag \\
    &\quad \times
    \begin{pmatrix}
        g_1^2 & g_1g_2 \\
        g_1g_2 & g_2^2
    \end{pmatrix}, \label{eq: the Flatte derived from EFT}
\end{align}

It is clear in Eq.~\eqref{eq: the Flatte derived from EFT} that the energy dependence of the Flatt\'{e} amplitude is common to all the coupled-channel components. The expansion of the denominator of each component in powers of $k$ is given by
\begin{align}
    f^{\rm F}_{ij}(E) &\propto \frac{1}{\frac{2E_{\rm BW}- ig_1^2p_0}{g_2^2}-\left(\frac{1}{\mu_2g_2^2}+i\frac{g_1^2}{2p_0g_2^2}\right)k^2 + {\cal O}(k^4) - ik}.
    \label{eq:fF11 effective range expansion}
\end{align}
In this way, all the components of the Flatt\'{e} amplitude are compatible with the effective range expansion, 
in contrast to the Contact amplitude. We thus determine the scattering length ($a_{\rm F}$) and effective range ($r_{\rm F}$) of the Flatt\'{e} amplitude as
\begin{align}
    a_{\rm F} &= \frac{g_2^2}{ig_1^2p_0 - 2E_{\rm BW}},
    \label{eq:Flatte scattering lengths}\\
    r_{\rm F} &= -\frac{2}{\mu_2g_2^2}-i\frac{g_1^2}{p_0g_2^2}.
    \label{eq:Flatte effective range}
\end{align}
The imaginary part of the effective range in Eq.~\eqref{eq:Flatte effective range} stems from the expansion of $p$ (momentum of channel 1) in terms of $k$. 
When the threshold energy difference $\Delta$ is large,  the momentum $p$ can be approximated by its threshold value $p_0$, as was done in Ref.~\cite{Baru:2004xg}. In this case, the imaginary part does not appear and $r_{\rm F}$ is obtained as a negative real value.

It was pointed out that the number of parameters in the Flatt\'{e} amplitude decreases near the threshold, due to the scaling behavior~\cite{Baru:2004xg}. From Eq.~\eqref{eq: the Flatte derived from EFT}, one finds that the Flatt\'{e} amplitude generally has three parameters, $E_{\rm BW},g_{1}$, and $ g_2$. To consider the scattering in the low-energy region near the threshold of channel 2, we keep the terms up to linear order in $k$ in the denominator, and $f^{\rm F}(E)$ takes the form
\begin{align}
    f^{\rm F}(E) &\simeq \frac{1}{\frac{\alpha}{R} p_0 - i\frac{1}{R}p_0 - ik}
    \begin{pmatrix}
        \frac{1}{R} & \sqrt{\frac{1}{R}} \\
        \sqrt{\frac{1}{R}} & 1
    \end{pmatrix},
    \label{eq:the Flatte amplitude up to first order in k which is represented by R and alpha}
\end{align} 
where we define the parameters $R$ and $\alpha$ as 
\begin{align}
    \frac{g_2^2}{g_1^2} = R, \quad
    \frac{2E_{\rm BW}}{g_1^2 p_0} = \alpha, \label{eq:the parameter alpha}
\end{align}
This gives the approximation of the Flatt\'{e} amplitude up to the first order of $k$. In the following, for simplicity, the approximated amplitude in Eq.~\eqref{eq:the Flatte amplitude up to first order in k which is represented by R and alpha} is also referred to as the Flatt\'{e} amplitude. Equation~\eqref{eq:the Flatte amplitude up to first order in k which is represented by R and alpha} shows that the Flatt\'{e} amplitude can be expressed only by two parameters $R$ and $\alpha$ near the threshold. In other words, three parameters of the Flatt\'{e} amplitude are reduced to two near threshold~\cite{Baru:2004xg}. As mentioned in Sec.~\ref{subsec: derivation of Contact}, the near-threshold two-channel scattering amplitude generally has three independent parameters. This indicates that some constraint is imposed on the Flatt\'{e} amplitude near the threshold. 

\subsection{
Comparison of two amplitudes
}\label{subsec: comparison C and F}
We have seen that there is some constraint in the Flatt\'{e} amplitude which reduces the number of independent parameters. To clarify this constraint, we focus on the determinant of the scattering amplitude matrix.
From the separable form of the residue in Eq.~\eqref{eq: the Flatte derived from EFT}, 
one finds that the rank of the matrix $f^{\rm F}(E)$ is one, and therefore its determinant vanishes:
\begin{align}
    \det\left[f^{\rm F}(E)\right] 
    &= 0. \label{eq: the determinant of Flatte amplitude}
\end{align}
This is caused by the same property of the potential $V_{\rm F}(E)$ we mentioned in Eq.~\eqref{eq: the effective potential for Flatte E}:
\begin{align}
    \det\left[V_{\rm F}(E)\right]  &= 0. \label{eq: the determinant of VF(E)}
\end{align}
As seen in Eq.~\eqref{eq: Flatte you lippman}, $\det\left[f(E)\right]=0 $ holds for the potential satisfying $\det\left[V(E)\right]=0$. Physically speaking, the vanishing of the determinant is due to the pole term nature of $V_{\rm F}(E)$ and $f^{\rm F}(E)$. It is known that the residue matrix of the pole term in the coupled-channel scattering should be rank one~\cite{Taylor}, and both $V_{\rm F}(E)$ and $f^{\rm F}(E)$ are given by a single pole term without any background scattering contribution. The condition~\eqref{eq: the determinant of Flatte amplitude} can be related to the constraint imposed on the Flatt\'{e} amplitude near the threshold. 

The Contact amplitude $f^{\rm C}(E)$ is considered as a general form of the two-channel scattering amplitude near the threshold. It is therefore expected that $f^{\rm C}(E)$ reduces to the Flatt\'{e} amplitude near the threshold by imposing the condition $\det\left[f^{\rm C}(E)\right]=0$. From Eq.~\eqref{eq:EFT amplitude}, the determinant of $f^{\rm C}(E)$ is given by 
\begin{align}
    \det\left[f^{\rm C}(E)\right] = -\frac{1}{\frac{1}{a_{12}^2} - \left(\frac{1}{a_{11}} + ip\right)\left(\frac{1}{a_{22}} + ik\right)}, \label{eq:detfEFT}
\end{align}
which is nonzero for finite $a_{11},a_{12}$, and $a_{22}$.
The condition $\det\left[f^{\rm C}(E)\right]=0$ is achieved by letting at least one of $a_{11},a_{12}$, and $a_{22}$ be zero. When we take the limit $a_{22}\rightarrow 0$ with $a_{12}\neq 0$ and $a_{11}\neq 0$, the Contact amplitude reduces to
\begin{align}
    f^{\rm C}(E) = \frac{1}{-\frac{1}{a_{11}}-ip(E)}
    \begin{pmatrix}
        1 & 0\\
        0 & 0
    \end{pmatrix}, \label{eq : EFT a22=0}
\end{align}
which represents the single-channel scattering in channel 1. This amplitude satisfies $\det[f^{\rm C}(E)]=0$ and does not have an inverse matrix, but is not the Flatt\'{e} amplitude. In the same way, taking only $a_{11}\rightarrow 0$ results in the elastic scattering amplitude in channel 2. The limit $a_{12}\rightarrow 0$ gives the the trivial amplitude with all components being zero. In this way, none of these limits can reproduce the coupled-channel Flatt\'{e} amplitude. Furthermore, setting two of the parameters zero with the remaining one being finite also results in the trivial amplitude. Therefore, in order to obtain a sensible coupled-channel amplitude, all the parameters must be simultaneously taken to be zero:
\begin{align}
    a_{11},a_{12},a_{22} \rightarrow 0. \label{eq: all a becomes 0}
\end{align}
In this case, the Contact amplitude is indefinite and the result depends on the ratios of $a_{ij}$.

In summary, we show that the difference between the Flatt\'{e} and Contact amplitudes can be characterized by the determinant of the amplitude, $\det[f^{\rm C}(E)]\neq 0$ and $\det[f^{\rm F}(E)]=0$. However, the Contact amplitude does not directly reduce to the Flatt\'{e} amplitude by imposing the condition $\det[f^{\rm C}(E)]= 0$ with the parameters $a_{11},a_{12}$, and $a_{22}$. In other words, $f^{\rm C}(E)$ is not a suitable form to study the connection with the Flatt\'{e} amplitude, although it represents the general form of the coupled-channel scattering amplitude.

\section{General amplitude}\label{sec : General amplitude}

In the previous section, we observe that the Flatt\'{e} amplitude follows the condition~$\det\left[f^{\rm F}(E)\right]=0$ and the Contact amplitude cannot directly be reduced to the Flatt\'{e} amplitude. In this section, we first discuss the condition
$\det\left[f(E)\right]=0$ in relation with the renormalization procedure in the Contact amplitude, and construct an expression of the scattering amplitude in which the condition $\det\left[f(E)\right]=0$ can be imposed by smoothly changing the parameters. We call this expression the General amplitude $f^{\rm G}$, and show that $f^{\rm G}$ reproduces both the Contact amplitude and the Flatt\'{e} amplitude in Sec.~\ref{sec:relation F and EFT}.
We extract the scattering length in the General amplitude and compare with that of the Flatt\'{e} amplitude in Sec.~\ref{subsec: slegth erange genral}. The decomposition of the General amplitude into the pole and background terms is discussed in Sec.~\ref{sec: pole plus bg}. 

\subsection{
Formulation
}\label{subsec: derivation of General}

To investigate the condition $\det\left[f^{\rm C}(E)\right]=0$ in more detail, we focus on the renormalization condition of the Contact amplitude. The first terms in the left hand side of the renormalization conditions~\eqref{eq:a_{11} kurikomi}-\eqref{eq:a_{12} kurikomi} contain $\det\left[V_{\rm C}\right]=c_{11}c_{22}-c_{12}^2$ in the denominator. In other words, $\det\left[V_{\rm C}\right]$ must be nonzero to adopt the renormalization condition~\eqref{eq:a_{11} kurikomi}-\eqref{eq:a_{12} kurikomi} that gives $a_{11},a_{12}$, and $a_{22}$. This explains the observation in the previous section that Eq.~\eqref{eq: all a becomes 0} is required
to achieve $\det\left[f^{\rm C}(E)\right]=0$. Therefore, to construct an amplitude which is smoothly connected to $\det\left[f^{\rm C}(E)\right]=0$, we need to modify the renormalization condition so that it is compatible with $\det\left[V_{\rm C}\right]=0$. For this purpose, we first consider the potential with two parameters $c_{22}$ and $x$
\begin{align}
    V_{\rm G}^{(0)} = 
    \begin{pmatrix}
        c_{22}x & c_{22}\sqrt{x} \\
        c_{22}\sqrt{x} & c_{22}
    \end{pmatrix},
     \label{eq: potential VEFT detV=0}
\end{align}
which is a special case of $V_{\rm C}$ with $c_{11}=c_{22}x$ and $c_{12} = c_{22}\sqrt{x}$, but at the same time the condition $\det[V_{\rm G}^{(0)}]=0$ is satisfied. Substituting $V_{\rm G}^{(0)}$ into the Lippmann-Schwinger equation~\eqref{eq: Flatte you lippman}, we obtain the T-matrix as 
\begin{align}
    T^{{\rm G}(0)}(E) &= \left[\frac{1}{c_{22}} + x\frac{\mu_1 \varLambda}{\pi^2} +\frac{\mu_2 \varLambda}{\pi^2}  + x\frac{\mu_1}{2\pi}ip  + \frac{\mu_2}{2\pi}ik\right]^{-1}\notag \\
    & \quad \times
    \begin{pmatrix}
        x  & \sqrt{x} \\
        \sqrt{x} &  1 
    \end{pmatrix}.\label{eq: Tmatrix when c11c22-c12^2=0}
\end{align}
We express the renormalized T-matrix with the physical quantities $A_{22},\epsilon$ as
\begin{align}
    T^{{\rm G}(0)}(E) &= \frac{1}{\frac{1}{A_{22}} + i\epsilon p + ik}
    \begin{pmatrix}
        \frac{2\pi}{\mu_1}\epsilon & \frac{2\pi}{\sqrt{\mu_1\mu_2}}\sqrt{\epsilon} \\
        \frac{2\pi}{\sqrt{\mu_1\mu_2}}\sqrt{\epsilon} & \frac{2\pi}{\mu_2}
    \end{pmatrix}. \label{eq: T-matrix represented by epsilon and A22}
\end{align}
This is achieved by the renormalization conditions 
\begin{align}
    x(\varLambda) &= \frac{\mu_2}{\mu_1} \epsilon, \label{eq: epsilon kurikomi c11c12c22} \\
    \frac{1}{c_{22}(\varLambda)} + x(\varLambda)\frac{\mu_1 \varLambda}{\pi^2} +\frac{\mu_2 \varLambda}{\pi^2} &= \frac{\mu_2}{2\pi}\frac{1}{A_{22}}. \label{eq: A22 kurikomi c11c12c22}
\end{align}
Taking the $\varLambda\to \infty$ limit, the renormalized scattering amplitude is obtained as
\begin{align}
    f^{{\rm G}(0)}(E) = \frac{1}{-\frac{1}{A_{22}} - i\epsilon p - ik}
    \begin{pmatrix}
        \epsilon & \sqrt{\epsilon} \\
        \sqrt{\epsilon} & 1
    \end{pmatrix}. \label{eq: the Flatte scattering amplitude represented by epsilon and A22}
\end{align}
In this way, we obtain the coupled-channel Contact amplitude with $\det [f(E)]=0$, by imposing the condition $\det [V]=0$ and modify the renormalization condition. On the other hand, comparing Eq.~\eqref{eq: the Flatte scattering amplitude represented by epsilon and A22} and Eq.~\eqref{eq:the Flatte amplitude up to first order in k which is represented by R and alpha}, we see that $f^{{\rm G}(0)}(E)$ is equivalent to the Flatt\'{e} amplitude up to the first order in $k$ with the identification
\begin{align}
    \epsilon &= \frac{1}{R}, \label{eq: ep R}\\
    A_{22} &= - \frac{R}{\alpha p_0}, \label{eq: A22 al}
\end{align}
in Eq.~\eqref{eq: the Flatte scattering amplitude represented by epsilon and A22}. In this way, we show that there is a well-defined limit of the Contact amplitude with $\det [f(E)]=0$, which is nothing but the Flatt\'{e} amplitude.
With these relations and Eq.~\eqref{eq:the parameter alpha}, the Flatt\'{e} scattering length $a_{\rm F}$ in Eq.~\eqref{eq:Flatte scattering lengths} can also be written by $A_{22}$ and $\epsilon$ as
\begin{align}
    a_{\rm F} &= \frac{1}{\frac{1}{A_{22}}+ i\epsilon p_0}.
    \label{eq:aFbyA22}
\end{align}

The remaining  task is to construct the scattering amplitude which is smoothly connected to $f^{{\rm G}(0)}(E)$. To this end, we propose the parametrization of the potential
 \begin{align}
     V_{\rm G} &= 
     \begin{pmatrix}
         c_{22}x & c_{22}\sqrt{x-y} \\
         c_{22}\sqrt{x-y} & c_{22}
     \end{pmatrix}, \label{eq: the general effective potential VG}
 \end{align}
with real bare parameters $c_{22},x$, and $y$. We require $x \geq y$ for the hermiticity of the Hamiltonian. The potential $V_{\rm G}$ is obtained by setting
\begin{align}
    c_{11} = c_{22}x, \quad
    c_{12} = c_{22}\sqrt{x-y}, \label{eq: c22 xy}
\end{align}
in $V_{\rm C}$, and because of three independent parameters. The determinant of $V_{\rm G}$ is obtained as
\begin{align}
    \det[V_{\rm G}] &= c^2_{22}y. \label{eq: detVg}
\end{align}
Therefore, the determinant of $V_{\rm G}$ vanishes with $y=0$ or $c_{22}=0$, but the latter is not of our interest because it yields the trivial scattering. 
With $y=0$, the potential becomes $V_{\rm G}^{(0)}$ in Eq.~\eqref{eq: potential VEFT detV=0}, which gives the Flatt\'{e} amplitude. Namely, the parametrization in Eq.~\eqref{eq: the general effective potential VG} allows one to control $\det[V_{\rm G}]$ by the parameter $y$.

Next, we derive the scattering amplitude using $V_{\rm G}$. Substituting $V_{\rm G}$ into Eq.~\eqref{eq: Flatte you lippman}, the T-matrix is obtained as 
\begin{widetext}
\begin{align}
    T^{\rm G}(E)
    &=
    \frac{1}{D(E)} 
    \begin{pmatrix}
        x 
        +
        c_{22}y
        \left(\frac{\mu_2\varLambda}{\pi^2} + \frac{\mu_2}{2\pi}ik
        \right)
        && \sqrt{x - y} \\
        \sqrt{x - y} && 
        1 +
        c_{22}y\left(\frac{\mu_1\varLambda}{\pi^2} + \frac{\mu_1}{2\pi}ip\right)
    \end{pmatrix},
    \label{eq: T bare parameter} \\
    D(E)&=
    \frac{1}{c_{22}} + x\frac{\mu_1 \varLambda}{\pi^2} +\frac{\mu_2 \varLambda}{\pi^2} 
    +c_{22}y\frac{\mu_1\mu_2\varLambda^2}{\pi^4} 
    + \left(x\frac{\mu_1}{2\pi}
    +c_{22}y\frac{\mu_1\mu_2\varLambda}{2\pi^3}\right)ip  
    + \left(\frac{\mu_2}{2\pi}
    +c_{22}y\frac{\mu_1\mu_2\varLambda}{2\pi^3}\right)ik 
    -c_{22}y\frac{\mu_1\mu_2}{4\pi^2}pk .
\end{align}
\end{widetext}
which recovers Eq.~\eqref{eq: Tmatrix when c11c22-c12^2=0} with $y=0$.
By introducing the $\varLambda$-dependence of $c_{22}(\varLambda),x(\varLambda)$, and $y(\varLambda)$ with the physical quantities $A_{22},\epsilon$, and $\gamma$ as
\begin{align}
    c_{22}(\varLambda) &= 2\pi^2A_{22}\gamma\left(\pi\mu_1 - 2\mu_1A_{22}\gamma\varLambda\right)\notag \\
    &\quad \times \left[ \left(2\mu_2A_{22}\gamma \varLambda - \pi \mu_2 \epsilon\right)\left(2\mu_1A_{22}\gamma\varLambda - \pi \mu_1\right)\right. \notag \\
    &\quad\left. - \pi^2\mu_1\mu_2(\epsilon - \gamma)\right]^{-1}, \label{eq: c11 kurikomi} \\
    x(\varLambda) &= \frac{2\mu_2A_{22}\gamma \varLambda - \pi \mu_2 \epsilon}{2\mu_1A_{22}\gamma\varLambda - \pi \mu_1}, \label{eq: R kurikomi} \\    
    y(\varLambda) &= \left[\left(2\mu_2A_{22}\gamma \varLambda - \pi \mu_2 \epsilon\right)\left(2\mu_1A_{22}\gamma\varLambda - \pi \mu_1\right)\right.\notag \\
    &\quad \left.- \pi^2\mu_1\mu_2(\epsilon - \gamma)\right]\left(\pi\mu_1 - 2\mu_1A_{22}\gamma\varLambda\right)^{-2}, \label{eq: x kurikomi}
\end{align}
we obtain the scattering amplitude $f^{\rm G}(E)$ in the $\varLambda \rightarrow \infty$ limit as
\begin{align}
    f^{\rm G}(E) 
    &= \frac{1}{-\frac{1}{A_{22}} - i\epsilon p - ik + A_{22} \gamma pk}\notag \\
    &\quad \times
    \begin{pmatrix}
        \epsilon + iA_{22}\gamma k & \sqrt{\epsilon-\gamma} \\
        \sqrt{\epsilon-\gamma} &  1 + iA_{22}\gamma p
    \end{pmatrix}.
    \label{eq:the general amplitude}
\end{align}
In the following, we call $f^{\rm G}(E)$ the General amplitude. When $\gamma\neq 0$, $f^{\rm G}(E)$ has an inverse matrix, which is given by
\begin{align}
    \left[f^{{\rm G}}(E)\right]^{-1} = 
    \begin{pmatrix}
        -\frac{1}{A_{22}}\frac{1}{\gamma} - ip & \frac{1}{A_{22}}\frac{\sqrt{\epsilon-\gamma}}{\gamma} \\
        \frac{1}{A_{22}}\frac{\sqrt{\epsilon-\gamma}}{\gamma} & -\frac{1}{A_{22}}\frac{\epsilon}{\gamma} - ik
    \end{pmatrix}. \label{eq: the scattring amplitude represented by A22, epsilon, gamma}
\end{align}
It is seen in this form that the physical parameters should satisfy $\epsilon>\gamma$ in order to be consistent with the optical theorem.

\subsection{
Relation with Contact and Flatt\'{e} amplitudes
}\label{sec:relation F and EFT}
\begin{figure}[tbp]
    \centering
    \includegraphics[width = 7cm, clip, 
    ]{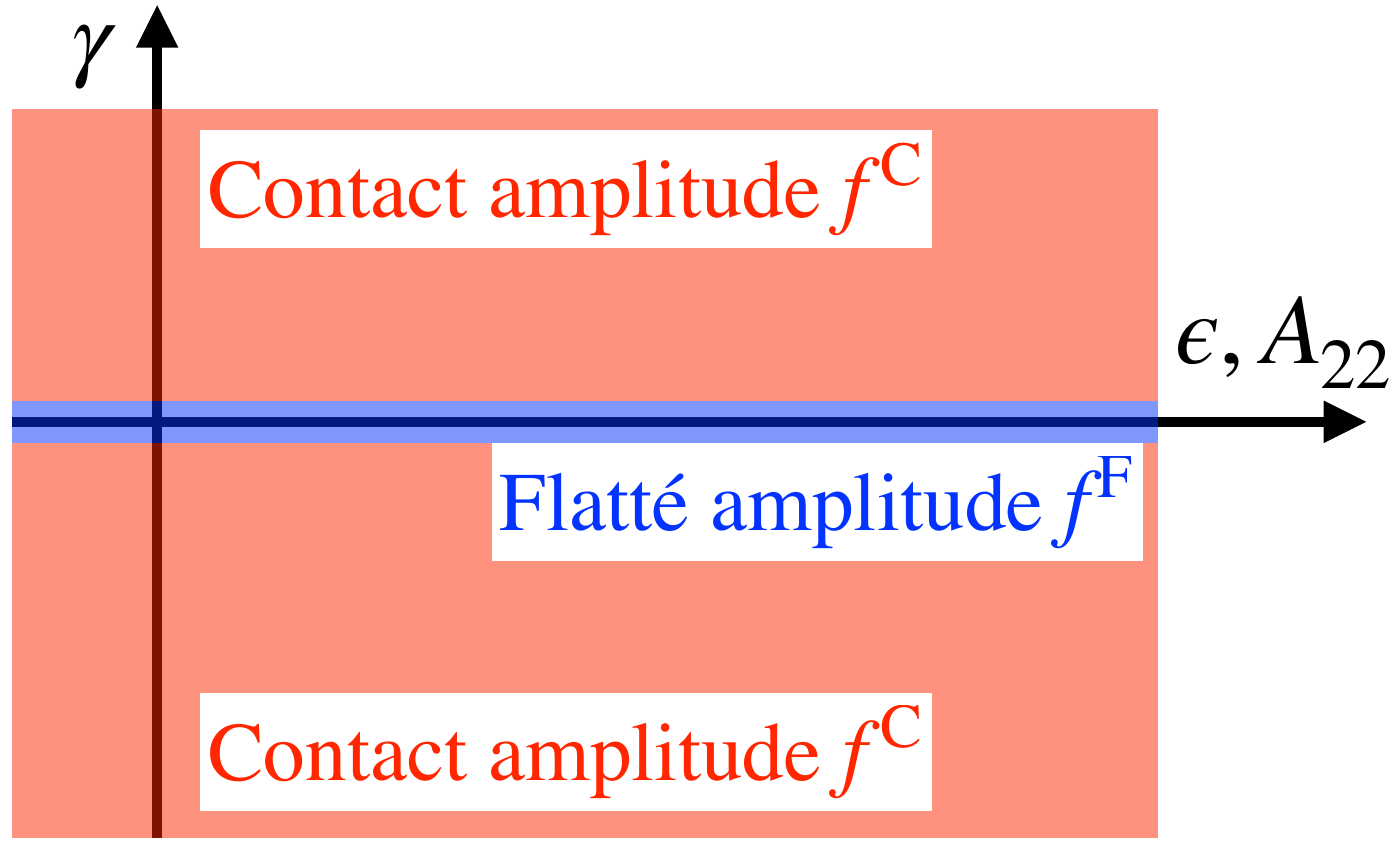}
    \caption{Schematic illustration of the parameter space of the General amplitude $f^{\rm G}$ in relation with the Contact amplitude $f^{\rm C}$ and Flatt\'{e} amplitude $f^{\rm F}$.
    }
    \label{fig: mosiki}
\end{figure}

In this section, we compare the General amplitude $f^{\rm G}(E)$ with the Contact amplitude $f^{\rm C}(E)$ and the Flatt\'{e} amplitude $f^{\rm F}(E)$ in terms of the parameters. First we show that $f^{\rm G}(E)$ contains both $f^{\rm C}(E)$ and $f^{\rm F}(E)$ in the parameter space. By comparing the inverse of the Contact amplitude Eq.~\eqref{eq:the invers matrix of the EFT amplitude} with that of the General amplitude~\eqref{eq: the scattring amplitude represented by A22, epsilon, gamma}, we find the relations 
\begin{align}
    a_{11} &= A_{22}\gamma, \label{eq: c p a11}\\
    a_{12} &= \frac{A_{22}\gamma}{\sqrt{\epsilon-\gamma}}, \label{eq: c p a12} \\
    a_{22} &= \frac{A_{22}\gamma}{\epsilon}. \label{eq: c p a22}
\end{align}
In this case, however, the condition $\gamma\neq 0$ is imposed in the General amplitude, as mentioned above. On the other hand, imposing the condition $\gamma=0$ on $f^{\rm G}(E)$ in Eq.~\eqref{eq:the general amplitude}, we obtain 
\begin{align}
    f^{\rm G}(E;A_{22}, \epsilon, \gamma=0) = \frac{1}{-\frac{1}{A_{22}} - i\epsilon p - ik }
    \begin{pmatrix}
        \epsilon & \sqrt{\epsilon} \\
        \sqrt{\epsilon} &  1
    \end{pmatrix}. \label{eq: Flatte fromfG}
\end{align}
Thus, we find that $f^{\rm G}(E)$ reduces to Eq.~\eqref{eq: the Flatte scattering amplitude represented by epsilon and A22} which is the Flatt\'{e} amplitude up to the first order in $k$. Therefore, we find that the General amplitude $f^{\rm G}(E)$ with $\gamma\neq0$ is the Contact amplitude $f^{\rm C}(E)$ and the General amplitude reduces to the Flatt\'{e} amplitude at $\gamma= 0$ (see Fig.~\ref{fig: mosiki}).

We are now in a position to revisit the relation between the Contact amplitude $f^{\rm C}(E)$ with the Flatt\'{e} amplitude $f^{\rm F}(E)$ using the parameters of the General amplitude $A_{22}, \epsilon$, and $ \gamma$. In Sec.~\ref{subsec: comparison C and F}, we have shown that the Contact amplitude cannot reduce to the Flatt\'{e} amplitude directly. First, from Eqs.~\eqref{eq: c p a11}-\eqref{eq: c p a22}, the condition  $\gamma=0$ indicates that all the parameters in the Contact amplitude vanish:
\begin{align}
    a_{11} &\rightarrow 0 \qquad (\gamma\rightarrow 0), \label{eq:a11 gamma 0}\\
    a_{12} &\rightarrow 0 \qquad (\gamma\rightarrow 0), \label{eq:a12 gamma 0}\\
    a_{22} &\rightarrow 0 \qquad (\gamma\rightarrow 0). \label{eq:a22 gamma 0}
\end{align}
This is in fact the condition discussed in the section \ref{subsec: comparison C and F} for the determinant of the Contact amplitude to vanish with the coupled-channel amplitude. Thanks to the relations~\eqref{eq: c p a11}-\eqref{eq: c p a22}, we can now establish the ratios of the $a_{ij}$ parameters in the $\gamma\to 0$ limit to obtain the Flatt\'{e} amplitude. For any value of $\gamma$, the following relations hold:
\begin{align}
    a_{11}\left(\frac{1}{a_{12}^2} - \frac{1}{a_{11}a_{22}}\right) &= - \frac{1}{A_{22}}, \label{eq: a11/a12 gama0} \\
    \frac{a_{11}}{a_{22}} &= \epsilon.\label{eq: a11/a22 gamma0} 
\end{align}
In addition, Eqs.~\eqref{eq: c p a11} and \eqref{eq: c p a12} shows that the relation
\begin{align}
    \frac{a_{11}}{a_{12}} &\rightarrow \sqrt{\epsilon} \qquad (\gamma\rightarrow 0), \label{eq: a12 gamma0}
\end{align}
holds in the limit of $\gamma\rightarrow 0$. Rewriting the Contact amplitude in Eq.~\eqref{eq:EFT amplitude} as
\begin{align}
    f^{\rm C}(E)&=  \frac{1}{a_{11}\left(\frac{1}{a_{12}^2} - \frac{1}{a_{11}a_{22}}\right) -ik - i\frac{a_{11}}{a_{22}}p  + a_{11}pk} \notag \\
    &\quad \times
    \begin{pmatrix}
        \frac{a_{11}}{a_{22}} + ia_{11}k & \frac{a_{11}}{a_{12}} \\
        \frac{a_{11}}{a_{12}} & 1 + ia_{11}p
    \end{pmatrix}, \label{eq: toyuu}
\end{align}
and sending all $a_{ij}$ to zero with keeping the ratios in Eqs.~\eqref{eq: a11/a12 gama0}-\eqref{eq: a12 gamma0}, we find that $f^{\rm C}(E)$ reduces to
\begin{align}
    \lim_{a_{11},a_{12},a_{22}\rightarrow 0}f^{\rm C}(E) &= \frac{1}{- \frac{1}{A_{22}} -ik - i\epsilon p}
    \begin{pmatrix}
        \epsilon & \sqrt{\epsilon} \\
        \sqrt{\epsilon} & 1
    \end{pmatrix},
\end{align}
which is the Flatt\'{e} amplitude in the form of $f_{\rm G}^{(0)}(E)$ in Eq.~\eqref{eq: the Flatte scattering amplitude represented by epsilon and A22}.

As mentioned in Sec.~\ref{subsec: comparison C and F}, the Contact amplitude cannot reproduce the Flatt\'{e} amplitude with finite parameters $a_{11},a_{12}$, and $a_{22}$. Here we show that the parametrization of the General amplitude with $A_{22}, \epsilon$, and $ \gamma$ is suitable to smoothly connect the Contact amplitude and the Flatt\'{e} amplitude. This is essential to examine the constraints in the cross sections from the Flatt\'{e} amplitude by the numerical analysis in the next section. At the same time, the General amplitude clarifies the conditions~\eqref{eq: a11/a12 gama0}, \eqref{eq: a11/a12 gama0}, and \eqref{eq: a12 gamma0} required to obtain the Flatt\'{e} amplitude from the Contact amplitude by taking the limit $a_{ij}\rightarrow 0$.

\subsection{
Scattering length
}\label{subsec: slegth erange genral}
In this section, we focus on the parameter $\gamma$ that characterizes the difference between the Contact amplitude and the Flatt\'{e} amplitude and discuss the effect of $\gamma$ on the scattering length in the General amplitude. First, we expand the General amplitude in $k$ (momentum of channel 2) and derive the scattering length of the General amplitude. Next, we show that the constant term in the denominator of the (1,1) component $f^{\rm G}_{11}$ is different from the value determined by the scattering length, except for the special case $\gamma=0$ corresponding to the Flatt\'{e} amplitude.

Expanding the denominator of the (2,2) component of the General amplitude $f^{\rm G}_{22}(E)$ in Eq.~\eqref{eq:the general amplitude}, we obtain 
\begin{align}
    f^{\rm G}_{22}(E) &= 
    \Biggl[-\frac{1}{A_{22}}\left(\frac{\frac{1}{A_{22}} + i\epsilon p_0}{\frac{1}{A_{22}} + i\gamma p_0}\right) - \frac{i\left(\epsilon-\gamma\right)}{2\left(1 + iA_{22}\gamma p_0\right)^2p_0}k^2 \notag \\
    &\quad - ik + {\cal O}(k^4)
    \Biggr]^{-1}. \label{eq:ERE of 22 component of the new amplitude}
\end{align}
Because this is consistent with the effective range expansion, the scattering length of the General amplitude is defined as
\begin{align}
    a_{\rm G} &= A_{22} \left(\frac{\frac{1}{A_{22}} + i\gamma p_0}{\frac{1}{A_{22}} + i\epsilon p_0}\right) \label{eq:the scattering length of the new amplitude} \\
    &= A_{22} \frac{\frac{1}{A_{22}^2}+\epsilon\gamma p_0^2}{\frac{1}{A_{22}^2} + \epsilon^2 p_0^2} - i\frac{\left(\epsilon - \gamma\right)p_0}{\frac{1}{A_{22}^2} + \epsilon^2 p_0^2}. \label{eq:Reag + Imag}
\end{align}
Thanks to the condition $\epsilon>\gamma$, the the requirement by the optical theorem $\text{Im} (a_{\rm G})<0$ is guaranteed in Eq.~\eqref{eq:Reag + Imag}.

On the other hand, the (1,1) component $f_{11}^{\rm G}(E)$ gives 
\begin{align}
    f^{\rm G}_{11}(E) &= 
    \frac{\epsilon^2}{\epsilon-\gamma}
    \left[-\frac{1}{A_{22}}\frac{\epsilon}{\epsilon-\gamma} - i\frac{\epsilon^2}{\epsilon-\gamma}p_0 \right. \notag \\
    &\quad \left.- \left(A_{22}\frac{\gamma}{\epsilon} + i\frac{1}{2p_0}\frac{\epsilon^2}{\epsilon-\gamma}\right)k^2 + {\cal O}(k^3) - ik\right]^{-1}.
    \label{eq:expansion of fG11 in k}
\end{align}
As in the case of the Contact amplitude in Eq.~\eqref{eq:ERE of 11 component of the EFT amplitude}, this expansion contains the odd power terms. In this way, we confirm that for $\gamma\neq 0$,  only the (2,2) component of the General amplitude can be written in the form of the effective range expansion, while the (1,1) component cannot. To discuss the relation between the scattering length of the General amplitude $a_{\rm G}$ and that of the Flatt\'{e} amplitude $a_{\rm F}$, we define $b_{\rm G}$ from the constant term of the denominator $f^{\rm G}_{11}(E)$ in Eq.~\eqref{eq:expansion of fG11 in k}:
\begin{align}
    f^{\rm G}_{11}(E) \propto \frac{1}{-\frac{1}{b_{\rm G}}-ik + {\cal O}(k^2)}, \label{eq: fG11 bG}
\end{align}
This gives $b_{\rm G}$ as
\begin{align}
    b_{\rm G} 
    &= \frac{\left(\epsilon - \gamma\right)\left(\frac{1}{A_{22}} - i\epsilon p_0\right)}{\epsilon \left(\frac{1}{A_{22}^2} + \epsilon^2 p_0^2\right)} \notag \\
    &= \frac{\left(\epsilon - \gamma\right)\frac{1}{A_{22}}}{\epsilon \left(\frac{1}{A_{22}^2} + \epsilon^2 p_0^2\right)} - i\frac{\left(\epsilon - \gamma\right)p_0}{\frac{1}{A_{22}^2} + \epsilon^2 p_0^2}. \label{eq: RebG + ImbG}
\end{align}
Comparison with Eq.~\eqref{eq:ERE of 22 component of the new amplitude} shows that the imaginary part of $b_{\rm G}$ is identical with that of $a_{\rm G}$. The real parts are, however, in general different from each other.

By setting $\gamma=0$, the real part of $b_{\rm G}$ also coincide with that of $a_{\rm G}$. In fact, from Eqs.~\eqref{eq:ERE of 22 component of the new amplitude} and \eqref{eq: RebG + ImbG},
 $a_{\rm G}$ and $b_{\rm G}$ with $\gamma=0$ are given by
\begin{align}
    a_{\rm G} &= \frac{1}{\frac{1}{A_{22}} + i\epsilon p_0} \quad (\gamma=0), \\
    b_{\rm G} &= \frac{1}{\frac{1}{A_{22}} + i\epsilon p_0} \quad (\gamma=0).
\end{align} 
This is natural because the General amplitude reduces to the Flatt\'{e} amplitude at $\gamma=0$. From Eq.~\eqref{eq:aFbyA22},
we find the relation
\begin{align}
    a_{\rm G} = b_{\rm G} = a_{\rm F} \qquad (\gamma=0), 
\end{align}
which shows that both $a_{\rm G}$ and $b_{\rm G}$ reduce to the scattering length of the Flatt\'{e} amplitude $a_{\rm F}$. In this way, we find that the scattering length of the Flatt\'{e} amplitude {$a_{\rm F}$} corresponds to the value in the special case with $\gamma=0$. In general, the constant term $b_{\rm G}$ defined in $f_{11}^{\rm G}(E)$ is different from the scattering length $a_{\rm G}$ given in $f_{11}^{\rm G}(E)$. 

\subsection{Pole and zero of amplitude
}\label{sec: pole plus bg}
In the previous section, it is found that the parameter $\gamma$ in the General amplitude characterizes the difference between the Contact amplitude and the Flatt\'{e} amplitude. In this section, we discuss the interpretation of this parameter $\gamma$, by studying the pole and zero of the amplitude. In the single-channel case, by truncating the effective range expansion in Eq.~\eqref{eq: 1 ERE} up to the first order in $k$, the amplitude has one pole at $k=k_{\rm p}=i/a$ determined by the scattering length $a$. The scattering amplitude is given only by the pole term:
\begin{align}
    f(k)=\frac{i}{k-k_{\rm p}} .
    \label{eq:fksingle}
\end{align}
In general, $f(k)$ has no zero unless we include a pole in the effective range expansion (Castillejo-Dalitz-Dyson zero~\cite{Castillejo:1955ed,Baru:2010ww,Kamiya:2017pcq}). For the comparison with the single-channel case, here we use the General amplitude up to the first order in $k$, by neglecting the $k$ dependence in the channel 1 momentum as $p=\sqrt{(\mu_1/\mu_2)k^2+2\mu_1\Delta}\simeq p_0=\sqrt{2\mu_1\Delta}$: 
\begin{align}
    f^{\rm G}(E) &= \frac{1}{-\frac{1}{A_{22}} - i\epsilon p_0 - ik + A_{22} \gamma p_0k} \notag \\
    &\quad \times
    \begin{pmatrix}
        \epsilon + iA_{22}\gamma k & \sqrt{\epsilon-\gamma} \\
        \sqrt{\epsilon-\gamma} &  1 + iA_{22}\gamma p_0
    \end{pmatrix}. \label{eq: fg up to 1 k}
\end{align}
This approximation is justified when the threshold energy difference $\Delta$ is sufficiently large.

First, we extract the pole term contribution from the General amplitude $f^{\rm G}(E)$. The General amplitude $f^{\rm G}(E)$ in Eq.~\eqref{eq: fg up to 1 k} has only one pole at $k=k_{\rm p}^{\rm G}$:
\begin{align}
    k^{\rm G}_{\rm p} &= \frac{\frac{1}{A_{22}} + i\epsilon p_0}{A_{22} \gamma p_0 - i} = \frac{i}{a_{\rm G}}, \label{eq: pole of fG}
\end{align}
which is common in all the components and determined by the scattering length $a_{\rm G}$, as in the case of the single-channel scattering. It can be seen that the pole appear near the threshold when the magnitude of the scattering length $|a_{\rm G}|$ is large. Let us rewrite each component of $f^{\rm G}(E)$ in Eq.~\eqref{eq: fg up to 1 k} with the pole momentum $k^{\rm G}_{\rm p}$. First, $f_{22}^{\rm G}(E)$ and $f^{\rm G}_{12}(E)$ components are given by
\begin{align}
    f_{22}^{\rm G}(E) &= \frac{i}{k-k_{\rm p}^{\rm G}}, \label{eq: fG22 represented by kp} \\
    f_{12}^{\rm G}(E) &= \frac{\sqrt{\epsilon-\gamma}}
    {1+iA_{22} \gamma p_0 }\frac{i}{k-k_{\rm p}^{\rm G}}, \label{eq: fG12 represented by kp}
\end{align}
which can be written only by the pole term proportional to $i/(k-k_{\rm p}^{\rm G})$. On the other hand, $f_{11}^{\rm G}(E)$ is written by
\begin{align}
    f_{11}^{\rm G}(E) = 
    \frac{\epsilon-\gamma}{(1+iA_{22} \gamma p_0 )^2} 
    \frac{i}{k-k_{\rm p}^{\rm G}} + \frac{A_{22}\gamma}{1+iA_{22} \gamma p_0 }, \label{eq: fG11 represented by kp}
\end{align}
which has a constant background term in addition to the pole term. In summary, $f^{\rm G}(E)$ can be decomposed into the pole term $f^{\rm G}_{\rm p}(E)$ and the background term $f^{\rm G}_{\rm p}(E)$: 
\begin{align}
    f^{\rm G}(E) &= f^{\rm G}_{\rm p}(E) + f^{\rm G}_{\rm bg},
    \label{eq: fG fp + fBg}\\
    f^{\rm G}_{\rm p}(E) &= \frac{i}{k-k_{\rm p}^{\rm G}}
    \begin{pmatrix}
        \frac{\epsilon-\gamma }{(1+iA_{22} \gamma p_0)^2} & \frac{\sqrt{\epsilon-\gamma}}{1+iA_{22} \gamma p_0 } \\
        \frac{\sqrt{\epsilon-\gamma}}{1+iA_{22} \gamma p_0 } & 1
    \end{pmatrix}, \\
     f^{\rm G}_{\rm bg} &= 
    \begin{pmatrix}
        \frac{A_{22}\gamma}{1+iA_{22} \gamma p_0 } & 0\\
        0 & 0
    \end{pmatrix}, \label{eq: bg term fG}
\end{align}
where one confirms $\det[f_{\rm p}^{\rm G}(E)]=0$ because of the rank one nature of the pole term. As mentioned above, up to the linear order in $k$, the single-channel scattering amplitude~\eqref{eq:fksingle} is given only by the pole term without the background contribution. In this sense, we can say that the appearance of the background term in the first order in momentum reflects the effect of the channel couplings. This is caused by the $k$ dependence in the numerator of the (1,1) component of $f^{\rm G}(E)$ in Eq.~\eqref{eq: fg up to 1 k}. Therefore, the existence of the background term $f^{\rm G}_{\rm bg}$ in the General amplitude $f^{\rm G}(E)$ is a property unique to the coupled channel scattering.

Imposing the condition $\gamma=0$ on Eq.~\eqref{eq: fG fp + fBg}, we obtain
\begin{align}
    f^{\rm G}(E;A_{22},\gamma=0,\epsilon) = \frac{i}{k-k_{\rm p}^{\rm F}}
    \begin{pmatrix}
        \epsilon & \sqrt{\epsilon} \\
        \sqrt{\epsilon} & 1
    \end{pmatrix}, \label{eq: Flatte fp}
\end{align}
where $k^{\rm F}_{\rm p}=i/A_{22}-\epsilon p_0$ is the pole of the Flatt\'{e} amplitude. When $\gamma=0$, the background term $f^{\rm G}_{\rm bg}$ disappears, because the Flatt\'{e} amplitude consists of the pure pole term. This indicates that the parameter $\gamma$ should be nonzero to have the background contribution. However, we see from Eq.~\eqref{eq: bg term fG} that the background term $f^{\rm G}_{\rm bg}$ depends on the parameter $A_{22}$ as well, so the magnitude of the background is not exclusively determined by $\gamma$. Furthermore, note that the pole position $k^{\rm G}_{\rm p}$ in Eq.~\eqref{eq: pole of fG} depends on $\gamma$ and therefore $\gamma$ plays an important role also in the pole term.

In general, the interference between the background term and the pole term can generate zeros of the scattering amplitude~\cite{Taylor}. Therefore, the (1,1) component of the General amplitude can have a zero point where $f^{\rm G}_{11}(E)=0$. From Eq.~\eqref{eq: fg up to 1 k}, the momentum $k^{\rm G}_{\rm zero}$ where the amplitude vanishes is given by
\begin{align}
    k^{\rm G}_{\rm zero} &= i \frac{1}{A_{22}}\frac{\epsilon} {\gamma}.
    \label{eq:zero point of the fG11 component interms of k}
\end{align}
Since $A_{22},\epsilon$, and $\gamma$ are the real parameters, we find that $k^{\rm G}_{\rm zero}$ is pure imaginary. Therefore, in terms of the energy variable, the zero point appears below the threshold. Furthermore, physical scattering below the threshold of channel 2 corresponds to the imaginary axis of the upper half of the complex $k$-plane ${\rm Im}(k)>0$. Therefore, the zero point of $f^{\rm G}_{11}(E)$ appears in the physical scattering region only if ${\rm Im}(k^{\rm G}_{\rm zero})>0$. On the other hand, the Flatt\'{e} amplitude (General amplitude with $\gamma=0$) has no zero point because there is no background contribution. This is reflected in Eq.~\eqref{eq:zero point of the fG11 component interms of k} which shows that $|k^{\rm G}_{\rm zero}|\rightarrow \infty$ and the zero point disappears in the limit $\gamma\to 0$. 

We have seen that the (1,1) component of the general amplitude has a zero point, which is absent in the Flatt\'{e} amplitude. Therefore, if the experimental data contains a zero point in the physical energy region ${\rm Im}(k)>0$ near the threshold, the analysis with the Flatt\'{e} amplitude would fail to reproduce the data.  In Sec.~\ref{sec: cs fg}, we quantitatively compare the Flatt\'{e} amplitude with the General amplitude in terms of the scattering cross section when $k^{\rm G}_{\rm zero}$ appears in the physical region near the threshold.

\section{Numerical analysis}\label{sec: cs fg}
In this section, by using the General amplitude, we examine the behavior of the total cross sections when a resonance pole of the scattering amplitude locates near the threshold of channel 2. Here, we investigate the behavior of the scattering cross section by  varying $\gamma$ with the scattering length $a_{\rm G}$ fixed. In this way, it is possible to compare the Flatt\'{e} amplitude $\gamma=0$ with the general case $\gamma \neq 0$, having the same amplitude at the threshold. First, in Sec.~\ref{subsec:henka A22}, we study the parameter regions of $A_{22}$ and $\epsilon$ when $\gamma$ varies for a fixed $a_{\rm G}$. Next, we summarize the actual values of the parameters used in the numerical calculation in Sec.~\ref{subsec:setup}. Finally, in Sec.~\ref{subsec:the cross section sigma11, sigma12}, we discuss the behavior of the scattering cross section numerically.

\subsection{
Parameter regions
}\label{subsec:henka A22}
The General amplitude $f^{\rm G}(E)$ in Eq.~\eqref{eq:the general amplitude} has three real parameters $A_{22},\epsilon$, and $\gamma$. We write the real and imaginary parts of the scattering length $a_{\rm G}$ in Eq.~\eqref{eq:Reag + Imag} as 
\begin{align}
    a_{\rm G} = \alpha + i\beta, \label{eq:alpha plus beta} 
\end{align}
using the real constants $\alpha$ and $\beta$, with the condition
$\beta<0$. Because $\alpha$ and $\beta$ are given by $A_{22},\epsilon$, and $\gamma$, when we fix $a_{\rm G}$, $A_{22}$ and $\epsilon$ are determined from $\alpha,\beta$, and $\gamma$. 
Since the relation between the parameters and the scattering length is not linear, we first discuss the behavior of the remaining two parameters $A_{22},\epsilon$ when the scattering length $a_{\rm G}$ is fixed.

Substituting the expression of the scattering length $a_{\rm G}$~\eqref{eq:the scattering length of the new amplitude} into Eq.~\eqref{eq:alpha plus beta}, we obtain
\begin{align}
    1 + iA_{22}\gamma p_0 &= \frac{\alpha}{A_{22}} - \epsilon \beta p_0 + i \left(\frac{\beta}{A_{22}}+\epsilon \alpha p_0\right) . \label{eq: aG = a +b}
\end{align}
The real and imaginary parts of Eq.~\eqref{eq: aG = a +b} yields the following two conditions:
\begin{align}
    1 &= \frac{\alpha}{A_{22}} - \epsilon \beta p_0, \label{eq:condition epsilon} \\
    A_{22}\gamma p_0 &= \frac{\beta}{A_{22}}+\epsilon \alpha p_0. \label{eq:condition A22}
\end{align}
Solving Eq.~\eqref{eq:condition epsilon} for $\epsilon$, we obtain 
\begin{align}
    \epsilon &= \frac{1}{\beta p_0}\left(\frac{\alpha}{A_{22}} - 1\right). \label{eq:epsilon represented by gamma}
\end{align}
We eliminate $\epsilon$ by substituting Eq.~\eqref{eq:epsilon represented by gamma} into Eq.~\eqref{eq:condition A22}. This gives a quadratic equation of $A_{22}$, whose two solutions are
\begin{align}
    A_{22}^{\pm} &= \frac{-\frac{\alpha}{\beta} \pm \sqrt{\left(\frac{\alpha}{\beta}\right)^2 + 4\gamma p_0 \left(\beta + \frac{\alpha^2}{\beta}\right)} }{2\gamma p_0}. \label{eq:A22 represented by gamma}
\end{align}
This gives the expression of the parameter $A_{22}$ in terms of $\alpha,\beta$, and $\gamma$. Corresponding $\epsilon$ is obtained from Eq.~\eqref{eq:epsilon represented by gamma} as
\begin{align}
    \epsilon^{\pm}(\alpha,\beta,\gamma) &= \frac{1}{\beta p_0}\left(\frac{\alpha}{A_{22}^{\pm}(\alpha,\beta,\gamma)} - 1\right), \label{eq: ep} 
\end{align}
In this way, there are two parameter sets $(A_{22}^{+},\epsilon^{+},\gamma)$ and $(A_{22}^{-},\epsilon^{-},\gamma)$ for a given scattering length $a_{\rm G}$.
To have a real parameter $A_{22}$, the quantity in the square root in Eq.~\eqref{eq:A22 represented by gamma} must be nonnegative, leading to the condition among parameters
\begin{align}
    \left(\frac{\alpha}{\beta}\right)^2 + 4\gamma p_0 \left(\beta + \frac{\alpha^2}{\beta}\right) & \geq 0. 
    \label{eq:conditionalphabetagamma}
\end{align}
Furthermore, considering $\beta <0$ required by the unitarity, we obtain the following condition for $\gamma$:
\begin{align}
    \frac{\left(\frac{\alpha}{\beta}\right)^2}{4 p_0 \left(|\beta| + \frac{\alpha^2}{|\beta|}\right)} & \geq \gamma. \label{eq: A22 repted abg}
\end{align}
This shows that the range of the parameter $\gamma$ has an upper bound   
\begin{align}
    \gamma_{\rm max} = \frac{\left(\frac{\alpha}{\beta}\right)^2}{4 p_0 \left(|\beta| + \frac{\alpha^2}{|\beta|}\right)}, \label{eq: max g}
\end{align}
to reproduce the given scattering length $a_{\rm G}$. We note that the inequality~\eqref{eq:conditionalphabetagamma} is saturated at $\gamma=\gamma_{\rm max}$ leading to $A_{22}^{+}=A_{22}^{-}$ and $\epsilon^{+}=\epsilon^{-}$. In other words, the two parameter sets $(A_{22}^{+},\epsilon^{+},\gamma)$ and $(A_{22}^{-},\epsilon^{-},\gamma)$ are continuously connected at $\gamma=\gamma_{\rm max}$.

\subsection{
Setup
}\label{subsec:setup}

To examine the behavior of $A_{22}$ and $\epsilon$ numerically, we consider the $\pi\pi$-$K\bar{K}$ system as an example. Since the threshold of $K\bar{K}$ ($\sim 990$ MeV) is far from that of $\pi\pi$ ($\sim 280$ MeV), we use the relativistic expression of the momentum
\begin{align}
    k &= \sqrt{\frac{(E+2m_{K})^2}{4} - m_K^2}, \label{eq: relative k}\\
    p &= \sqrt{\frac{(E+2m_{K})^2}{4} - m_{\pi}^2} \label{eq: relative p}.
\end{align}
In this case, the momentum $p$ at the threshold of channel 2 is evaluated as $p_0=\sqrt{m_K^2-m_\pi^2}$. 
The hadron masses are taken from PDG~\cite{ParticleDataGroup:2022pth}. 

\begin{figure}[tbp]
    \centering
    \includegraphics[width = 7cm, clip]{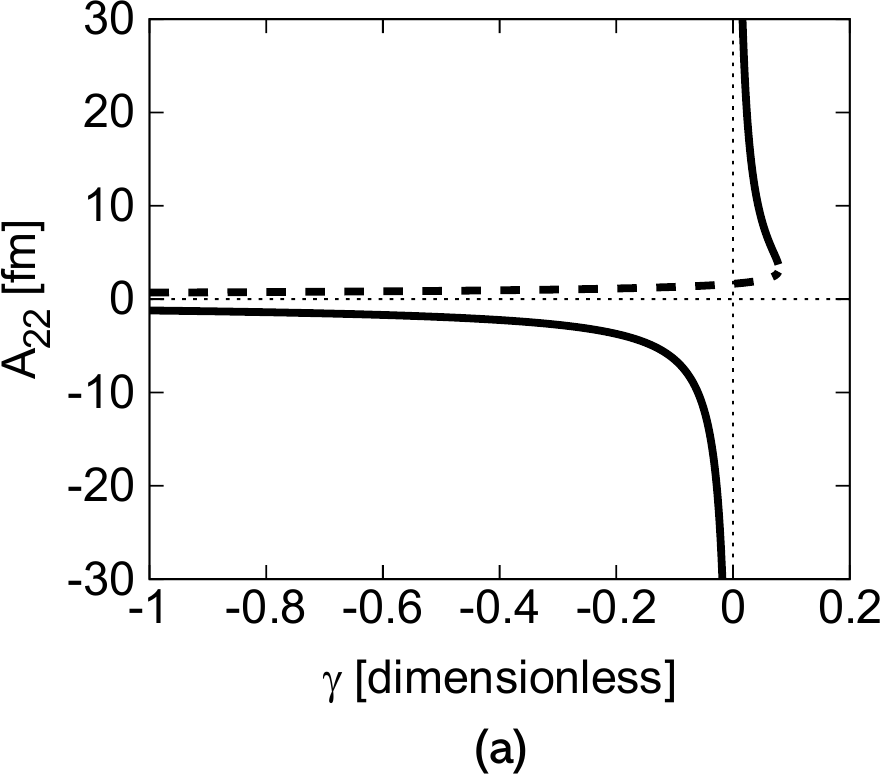}
    \includegraphics[width = 7cm, clip]{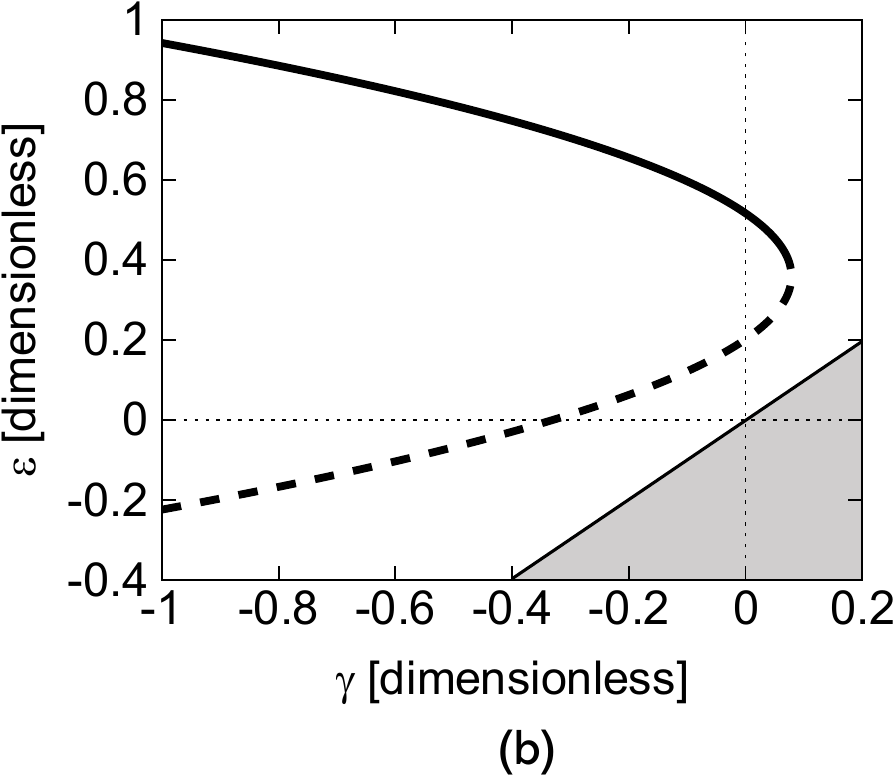}
    \caption{
    Parameters $A_{22}$ (a) and $\epsilon$ (b) as functions of $\gamma$ with the scattering length $a_{\rm G}=+1.0-i0.8\ {\rm fm}$. The solid (dashed) lines represent $A_{22}^+$ and $\epsilon^+$ ($A_{22}^-$ and $\epsilon^-$).
    The shaded area is excluded by the condition $\epsilon>\gamma$ by the optical theorem. 
    }
    \label{fig:A22plus1}
\end{figure}

First, we fix the scattering length as
\begin{align}
    a_{\rm G} &= +1.0 - i0.8 \quad {\rm fm},  
    \label{eq:aGpositive}
\end{align}
which is a typical value found in the experimental analysis~\cite{KLOE:2005jxf}. By Eq.~\eqref{eq: pole of fG}, the pole of $f_0(980)$ is estimated as
\begin{align}
    k^{\rm G}_{\rm p} &= -0.095+i0.125\ {\rm GeV}, \\
    E^{\rm G}_{\rm p} &= \frac{[k^{\rm G}_{\rm p}]^2}{m_K} = 
    -0.014-i0.048\ {\rm GeV} .
\end{align}
Because the imaginary part of the eigenmomentum is positive, this pole represent the quasibound state~\cite{Nishibuchi:2023acl}. 
Note however that the pole position in Eq.~\eqref{eq: pole of fG} is obtained by approximating $p\sim p_0$, and the exact pole of the amplitude calculated with $p(k)$ depends not only on $a_{\rm G}$ but also $\gamma$. 
In the following calculations, we have checked that the exact pole is found within a few MeV region from the above values.
Also, the upper bound $\gamma_{\rm max}$ in Eq.~\eqref{eq: max g} is given by
\begin{align}
    \gamma_{\rm max} = 0.079.
    \label{eq:gammamaxnumerical}
\end{align}
Numerical plots of $A_{22}^{+}$ (solid line) and $A^{-}_{22}$ (dashed line) for $-1.0\leq \gamma\leq \gamma_{\rm max}$ are shown in Fig.~\ref{fig:A22plus1}(a). As expected, two lines coincide with each other at $\gamma=\gamma_{\rm max}$. From Fig.~\ref{fig:A22plus1}(a), we find that while $A_{22}^{-}$ has a finite value at $\gamma=0$, $A_{22}^+$ diverges as $A_{22}^+\to \pm\infty$ with $\gamma\to 0^{\pm}$. That is, for $a_{\rm G} = +1.0 - i0.8 \ [\rm fm]$, the Flatt\'{e} amplitude is obtained by adopting the solution $A_{22}^{-}$ and setting $\gamma \rightarrow 0$. Next, the plots of $\epsilon^\pm$ as functions of $\gamma$ are shown in Fig.~\ref{fig:A22plus1}(b). The solid line represents $\epsilon^+$ and the dashed line $\epsilon^{-}$. From Fig.~\ref{fig:A22plus1}(b), we can see that $\epsilon^+$ is always positive in this range of $\gamma$, while the sign of $\epsilon^-$ changes around $\gamma\sim -0.4$. This property will be discussed in relation with the zero point of the scattering cross section. It can be checked that the condition $\epsilon>\gamma$ is satisfied for both solutions in this parameter region.

\begin{figure}[tbp]
    \centering
    \includegraphics[width = 7cm, clip]{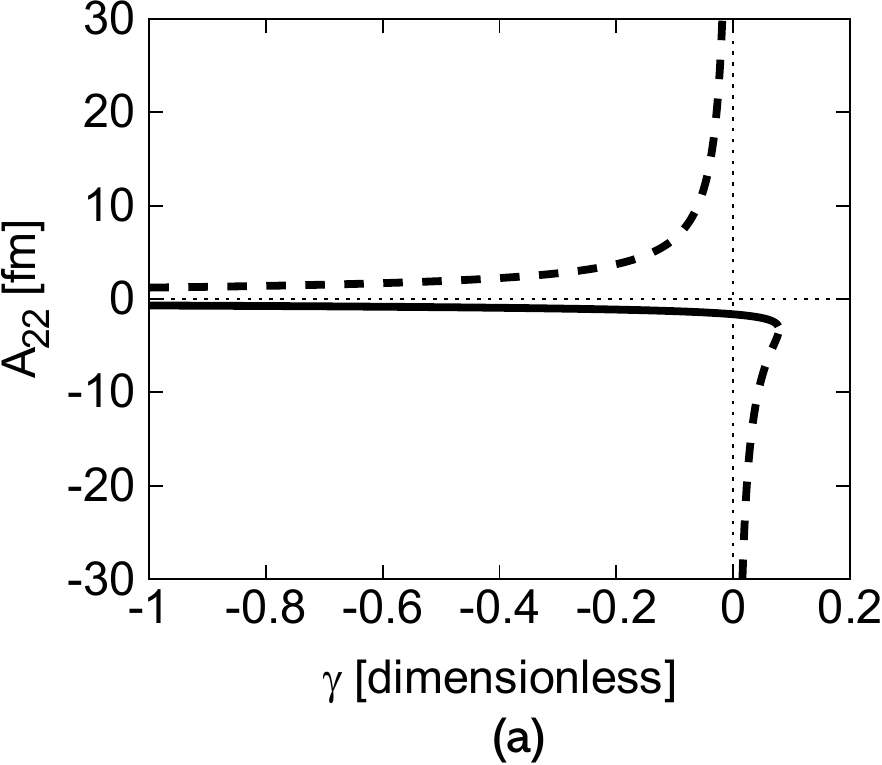}
    \includegraphics[width = 7cm, clip]{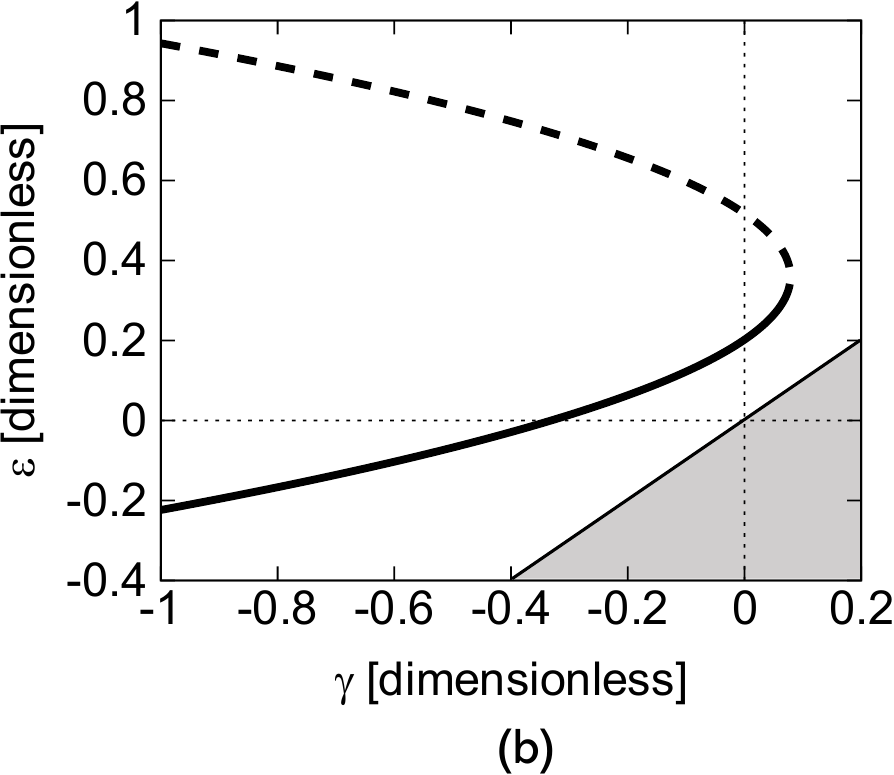}
    \caption{
    Same as Fig.~\ref{fig:A22plus1} but with $a_{\rm G}=-1.0-i0.8\ {\rm fm}$.}
    \label{fig:epsilon1}
\end{figure}

It is instructive to consider the case with the scattering length having the opposite sign of the real part from Eq.~\eqref{eq:aGpositive}. For this purpose, we also examine
\begin{align}
    a_{\rm G} = -1.0 - i0.8\quad {\rm fm}.
\end{align}
Corresponding pole position of $f_0(980)$ in Eq.~\eqref{eq: pole of fG} is found to be
\begin{align}
    k^{\rm G}_{\rm p} &= -0.095-i0.125\ {\rm GeV}, \\
    E^{\rm G}_{\rm p} &= -0.014+i0.048\ {\rm GeV}.
\end{align}
Negative imaginary part of the eigenmomentum suggests that this pole would be a virtual state in the absence of the coupling to the $\pi\pi$ channel. 
According to Ref.~\cite{Nishibuchi:2023acl} we call this the quasivirtual state. Because $\gamma_{\rm max}$ in Eq.~\eqref{eq: max g} is invariant under $\alpha\to -\alpha$, the value in Eq.~\eqref{eq:gammamaxnumerical} still holds in the present case. We plot $A_{22}$ and $\epsilon$ as functions of $\gamma$ in Fig.~\ref{fig:epsilon1}. Comparing $A_{22}^\pm$ in Fig.~\ref{fig:A22plus1}(a) with those in Fig.~\ref{fig:epsilon1}(a) we notice that there is some symmetry. In fact, from  Eq.~\eqref{eq:A22 represented by gamma}, we find the relation $A_{22}^{\pm}(-\alpha)=-A_{22}^{\mp}(\alpha)$ under the sign flip of the real part of the scattering length, which explains the behaviors in Fig.~\ref{fig:A22plus1}(a) and Fig.~\ref{fig:epsilon1}(a). This suggests that the Flatt\'{e} amplitude is obtained by adopting $A_{22}^{+}$ and setting $\gamma \rightarrow 0$ for $a_{\rm G}=-1.0-i0.8\ {\rm fm}$. The behaviors of $\epsilon^\pm$ in  Figs~\ref{fig:A22plus1}(b) and~\ref{fig:epsilon1}(b) can also be explained by the relation $\epsilon^{\pm}(-\alpha)=\epsilon^{\mp}(\alpha)$ which follows from Eq.~\eqref{eq: ep} and the property of $A_{22}^\pm$.

\subsection{
Total cross sections
}\label{subsec:the cross section sigma11, sigma12}

In the previous section, we established the behaviors of $A_{22}$ and $\epsilon$ when the scattering length $a_{\rm G}$ is fixed and $\gamma$ is varied. In this section, we discuss the behavior of the scattering cross section with representative values of the parameters. In the near threshold region where the $s$-wave contribution dominates, the total cross section $\sigma_{ij}(E)$ is given by  
\begin{align}
    \sigma_{ij}(E) &= 4\pi\frac{p_i}{p_j}|f_{ij}(E)|^2. \label{eq:relation between f and sigma}
\end{align}
Because the cross section is proportional to $|f_{ji}(E)|^2$, in this section, we use 
\begin{align}
    \sigma_{ij}^{\rm N}(E) \equiv |f_{ij}(E)|^2 / |f_{ij}(0)|^2, \label{eq: sigma kikakuka}
\end{align}
which is normalized at the threshold.

Before performing the numerical analysis, let us examine the analytic properties of $\sigma_{ij}^{\rm N}(E)$. From Eq.~\eqref{eq: fg up to 1 k}, the (1,1) component of the cross section $\sigma_{11}^{\rm N}(E)$ is given by
\begin{align}
    \sigma_{11}^{\rm N}(E) &= \left|\frac{1+i\frac{A_{22}\gamma}{\epsilon}k}{1+ia_{\rm G}k}\right|^2,
    \label{eq:sigmaG11}
\end{align}
which depends on three parameters $A_{22},\epsilon$, and $\gamma$. Thus, even if we fix the scattering length $a_{\rm G}$, the cross section still depends on the parameter $\gamma$. In contrast, the (1,2) and (2,2) components of the normalized cross section $\sigma^{\rm N}(E)$ are given by
\begin{align}
    \sigma^{\rm N}_{12}(E) = \sigma^{\rm N}_{22}(E) = \left|\frac{1}{1+ia_{\rm G}k}\right|^2. \label{eq: sigmaG12}
\end{align}
which depends exclusively on the scattering length $a_{\rm G}$. In other words, they are independent of $\gamma$ when $a_{\rm G}$ is fixed. In this way, the $\gamma$ dependence of the normalized cross section exists only in the (1,1) component, because only $f^{\rm G}_{11}(E)$ contains the background term as discussed in Sec.~\ref{sec: pole plus bg}.

Now we calculate the normalized cross sections near the threshold of channel 2 by varying $\gamma$ with the scattering length $a_{\rm G}$ being fixed. We focus on $\sigma_{11}^{\rm N}(E)$ that has $\gamma$ dependence for a fixed $a_{\rm G}$. First, we set
\begin{align}
    a_{\rm G} = +1.0 - i0.8 \quad {\rm fm} \notag
\end{align}
Because $\gamma$ should be smaller than $\gamma_{\rm max}$, we choose $\gamma=0.07,\ 0.00,\ -0.01,\ -10.0$ as representative values. The corresponding parameters  $A_{22}^{\pm}$ and $ \epsilon^{\pm}$ are shown in Table~\ref{table1}. The Flatt\'{e} amplitude corresponds to the solution with $A_{22}^{-}$ and $ \epsilon^{-}$ at $\gamma=0$. 

\begin{table}
\caption{
Parameters $A^{\pm}_{22}$ and $\epsilon^\pm$ for $a_{\rm G}=+1.0-i0.8$ fm.}
\label{table1}
\begin{ruledtabular}
\begin{tabular}{ccccc}
$\gamma$ & $A_{22}^{+}$\ (\rm fm) & $\epsilon^{+}$ & $A_{22}^{-}\ (\rm fm)$& $\epsilon^{-}$ \\ \hline
        $+0.07$& $+4.95$ & $+0.41$ & $+2.45$ & $+0.31$  \\
        $0.00$ & - & $+0.52$ & $+1.64$ & $+0.20$ \\
        $-0.01$ & $-53.4$ & $+0.53$ & $+1.59$ & $+0.19$  \\
        $-10.0$ & $-0.32$ & $+2.14$ & $+0.27$ & $-1.42$ \\
\end{tabular}
\end{ruledtabular}
\end{table}

\begin{figure}[tbp]
    \centering
    \includegraphics[width = 8cm, clip]{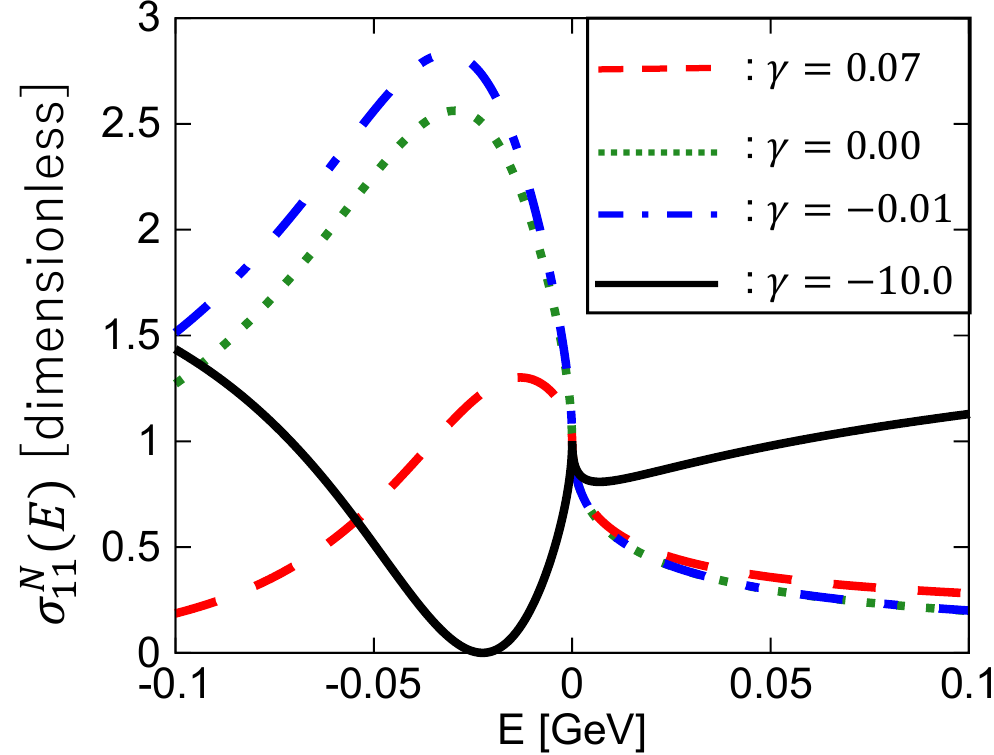}
    \caption{
    The (1,1) component of the normalized cross sections $\sigma_{11}^{\rm N}(E)=|f^{\rm G}_{11}(E)|^2/|f^{\rm G}_{11}(0)|^2$ as functions of the energy $E$. The scattering length is fixed as $a_{\rm G}=+1.0-i1.0\ {\rm fm}$ and the solution with $A_{22}^{-}$ and $\epsilon^{-}$ is chosen. The parameter is $\gamma=0.07$ (dashed line), $\gamma=0.00$ (dotted line), $\gamma=-0.01$ (dash-dotted line), and $\gamma=-10.0$ (solid line).
    }
    \label{fig:sigmaG11 A22-, epsilon-, aG=+1.0-i1.0}
\end{figure}

Choosing $A_{22}^{-}$ and $ \epsilon^{-}$, we plot the (1,1) component of the cross section $\sigma_{11}^{\rm N}(E)$ for four values of $\gamma$ in Fig.~\ref{fig:sigmaG11 A22-, epsilon-, aG=+1.0-i1.0}. The dotted line ($\gamma=0$) corresponds to the cross section by the Flatt\'{e} amplitude which shows the peak structure below the threshold. This is because in this amplitude there is a quasibound state below the threshold, as discussed above. The peak locates around $E\sim -0.03$ GeV, which is shifted from the real part of the pole energy $\sim -0.01$ GeV due to the threshold effect. 
When $\gamma$ is increased from zero, the peak structure remains, but the size of the peak becomes smaller than that with $\gamma=0$, as seen by the dashed line $(\gamma=0.07)$. The peak position moves toward the threshold. On the other hand, the peak becomes larger when $\gamma$ is slightly decreased from zero in the negative direction ($\gamma=-0.01$, dash-dotted line). 
In this way, the cross sections with small $|\gamma|$ have a peak structure, for which the Flatt\'{e} amplitude can be used to fit the data. However, if we fit the peaks of, for instance, the dashed or dash-dotted lines in Fig.~\ref{fig:sigmaG11 A22-, epsilon-, aG=+1.0-i1.0} using the Flatt\'{e} amplitude, the scattering length would be different from the exact value $+1.0-i0.8\ {\rm fm}$. Because the value of $\gamma$ is not known in advance in the experimental data, the scattering length extracted by the Flatt\'{e} amplitude might be deviated from the exact value.

When we adopt a large and negative $\gamma=-10.0$ (solid line in Fig.~\ref{fig:sigmaG11 A22-, epsilon-, aG=+1.0-i1.0}), the cross section $\sigma_{11}^{\rm N}(E)$ no longer shows the peak structure but exhibits a dip structure instead,  in sharp contrast to the other cases. In the present case, at the bottom of this dip, the cross section $\sigma_{11}^{\rm N}(E)$ becomes exactly zero, meaning that the scattering amplitude vanishes at that energy. This is the manifestation of the Castillejo-Dalitz-Dyson zero~\cite{Castillejo:1955ed,Baru:2010ww,Kamiya:2017pcq}, caused by the interference between the pole and background, as discussed in Sec.~\ref{sec: pole plus bg}. From Eq.~\eqref{eq:zero point of the fG11 component interms of k}, the zero point of the (1,1) component $f^{\rm G}_{11}(E)$ of the General amplitude is calculated as
\begin{align}
    k^{\rm G}_{\rm zero} &= i0.10\ {\rm GeV},\\
    E^{\rm G}_{\rm zero} &= 
    \frac{[k^{\rm G}_{\rm zero}]^2}{m_K} = 
    -0.02\ {\rm GeV}.\label{eq: Ep1}
\end{align}
In fact, the zero point of the cross section indeed occurs in Fig.~\ref{fig:sigmaG11 A22-, epsilon-, aG=+1.0-i1.0} at the value of Eq.~\eqref{eq: Ep1}. When the cross section shows the dip structure as the solid line ($\gamma=-10.0$) in Fig.~\ref{fig:sigmaG11 A22-, epsilon-, aG=+1.0-i1.0}, the Flatt\'{e} amplitude does not work. As shown in Sec.~\ref{sec: pole plus bg}, there is no zero point in the Flatt\'{e} amplitude, and thus no zero of the cross section occurs. In addition, the typical Flatt\'{e} amplitude exhibits either the peak structure or the threshold cusp structure~\cite{Baru:2004xg}; the former corresponds to the dotted line in Fig.~\ref{fig:sigmaG11 A22-, epsilon-, aG=+1.0-i1.0} and the latter to the cross section with $a_{\rm G} = -1.0-i0.8\ {\rm fm}$ discussed below. These structures are clearly different from the dip structure shown by the solid line ($\gamma=-10.0$) in Fig.~\ref{fig:sigmaG11 A22-, epsilon-, aG=+1.0-i1.0}. In contrast to the typical cross sections, the solid line in Fig.~\ref{fig:sigmaG11 A22-, epsilon-, aG=+1.0-i1.0} does not show the peak nor cusp, even though the pole is near the threshold. This result indicates that the analysis of the exotic hadrons requires a detailed study of the behavior of the scattering cross section near the threshold, rather than focusing only on peaks and cusp (see also Refs.~\cite{Dong:2020hxe,Baru:2024ptl}).

In general, the zero point appears in the physical region when $\epsilon/(A_{22}\gamma)>0$. From Fig.~\ref{fig:A22plus1}, this corresponds to $\gamma\lesssim -0.4\ (A_{22}^{-}>0, \epsilon^{-} <0, \gamma<0)$ or $0<\gamma<\gamma_{\rm max}\ (A^{-}_{22}>0, \epsilon^{-} >0, \gamma>0)$. Therefore, the dashed line ($\gamma=0.07$) in Fig.~\ref{fig:sigmaG11 A22-, epsilon-, aG=+1.0-i1.0} also has a zero point somewhere in the $E<0$ region, but the dash-dotted line ($\gamma=-0.01$) does not. As discussed in Eq.~\eqref{eq:zero point of the fG11 component interms of k}, when $\gamma$ is small, the corresponding momentum becomes large and the zero point appears far from the threshold. In fact, the zero point energy for $\gamma=0.07$ is $E=-0.25$ GeV which locates outside of the figure.

\begin{figure}[tbp]
    \centering
    \includegraphics[width = 8cm, clip]{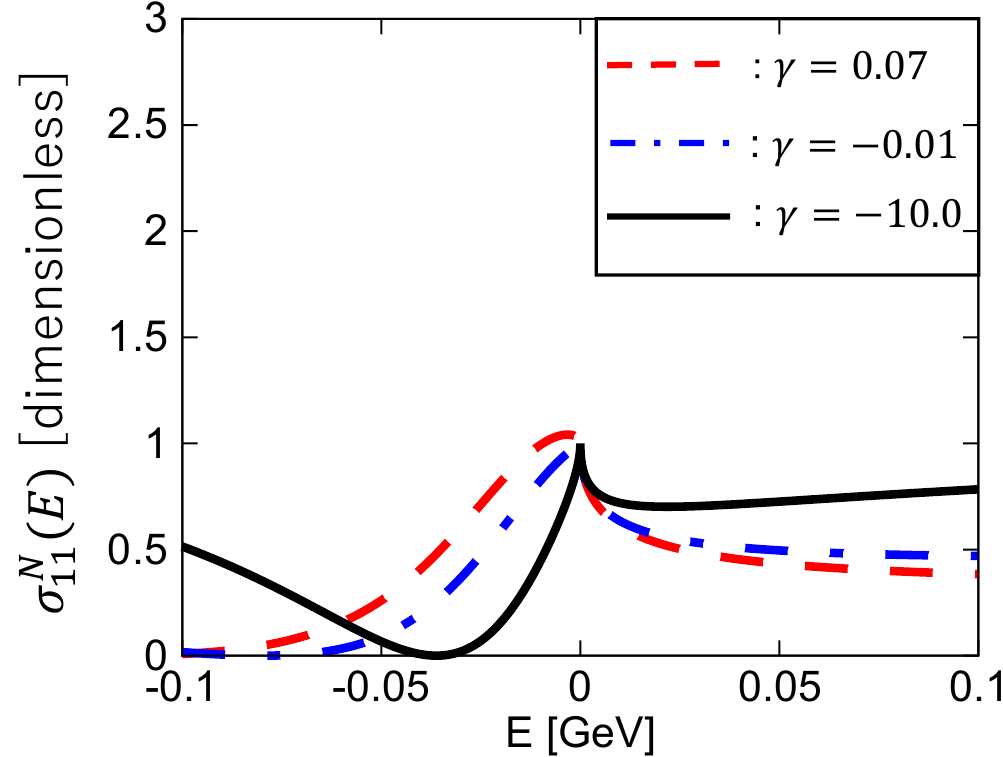}
    \caption{
    Same as Fig.~\ref{fig:A22plus1} but with $A_{22}^{+}$ and $\epsilon^{+}$.
    }
    \label{fig:sigmaG11 A22+, epsilon+, aG=+1.0-i1.0}
\end{figure}

The results with $A_{22}^{+}$ and $\epsilon^{+}$ are shown in  Fig.~\ref{fig:sigmaG11 A22+, epsilon+, aG=+1.0-i1.0}. The case with $\gamma=0$ is not shown because $A_{22}^{+}$ diverges. As mentioned in Sec.~\ref{subsec:henka A22}, $A_{22}^{+}$ and $\epsilon^{+}$ are continuously connected with $A_{22}^{-}$ and $\epsilon^{-}$ at $\gamma = \gamma_{\rm max}$, so the dashed line with $\gamma \sim \gamma_{\rm max}$ in Fig.~\ref{fig:sigmaG11 A22+, epsilon+, aG=+1.0-i1.0} shows a smaller peak structure than the dashed line in Fig.~\ref{fig:sigmaG11 A22-, epsilon-, aG=+1.0-i1.0}. In Fig.~\ref{fig:sigmaG11 A22+, epsilon+, aG=+1.0-i1.0}, when $\gamma=-0.01$ (dash-dotted line) the peak structure of the cross section is almost invisible. When $\gamma$ is further decreased down to $\gamma=-10.0$ (solid line), the zero point appears in the physical region below the threshold, generating a sharp dip structure. It can be seen from Fig.~\ref{fig:A22plus1} that the zero point is always in the physical region irrespective to the value of $\gamma$, since the relations $\epsilon^{+}>0$ and $ A_{22}^{+}\gamma>0$ always hold. The results in Fig.~\ref{fig:sigmaG11 A22+, epsilon+, aG=+1.0-i1.0} show that if the zero point is far enough from the threshold with small $|\gamma|$, the peak structure is preserved, but if the zero point appears near the threshold with large $|\gamma|$, the dip structure is produced.
Note also that the solid line in Fig.~\ref{fig:sigmaG11 A22+, epsilon+, aG=+1.0-i1.0} shows a cusp at the threshold, even though the amplitude has a quasibound pole. This is again peculiar feature in the General amplitude, which cannot be reproduced by the Flatt\'{e} amplitude.

\begin{figure}[tbp]
    \centering
    \includegraphics[width = 8cm, clip]{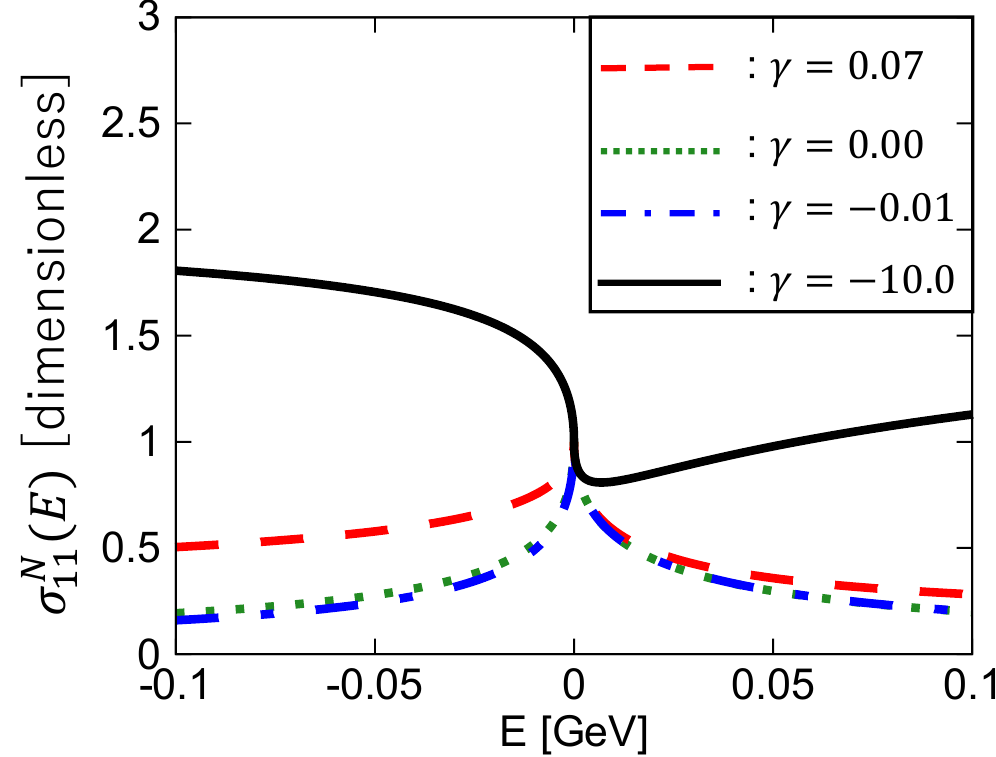}
    \caption{
    Same as Fig.~\ref{fig:A22plus1} but the scattering length is fixed as  $a_{\rm G}=-1.0-i0.8\ {\rm fm}$ with $A_{22}^{+}$ and $\epsilon^{+}$.
    }
    \label{fig:sigmaG11 cusp A22+, epsilon+, aG=-1.0-i1.0}
\end{figure}
\begin{table}
\caption{
Parameters $A^{\pm}_{22}$ and $\epsilon^\pm$ for $a_{\rm G}=-1.0-i0.8$ fm.}
\label{table2}
\begin{ruledtabular}
\begin{tabular}{ccccc}
$\gamma$ & $A_{22}^{+}$\ (\rm fm) & $\epsilon^{+}$ & $A_{22}^{-}\ (\rm fm)$& $\epsilon^{-}$ \\ \hline
        $+0.07$& $-2.45$ & $+0.31$ & $-4.95$ & $+0.41$  \\
        $0.00$ & $-1.64$ & $+0.20$ & - & $+0.52$ \\
        $-0.01$ & $-1.59$ & $+0.19$ & $+53.4$ & $+0.53$  \\
        $-10.0$ & $-0.27$ & $-1.42$ & $-0.32$ & $+2.14$ \\
\end{tabular}
\end{ruledtabular}
\end{table}

Next, we discuss the cross sections with
the scattering length 
\begin{align}
    a_{\rm G} = -1.0 - i0.8\ {\rm fm}.
\end{align}
The corresponding parameters $A_{22}^{\pm}$ and $\epsilon^\pm$ for $\gamma=0.07,\ 0.00,\ -0.01,\ -10.0$ are shown in Table.~\ref{table2}. Choosing $A_{22}^{+}$ and $\epsilon^+$, we plot the normalized cross sections $\sigma_{11}^{\rm N}(E)$ with $\gamma=0.07,\ 0.00,\ -0.01,\ -10.0$ in Fig.~\ref{fig:sigmaG11 cusp A22+, epsilon+, aG=-1.0-i1.0}. With $\gamma=0$ (Flatt\'{e} amplitude), the cross section shows a threshold cusp structure. Because the pole of the quasivirtual state is not on the most adjacent Riemann sheet to the physical scattering, it does not create a peak structure of the cross section. The existence of the quasivirtual pole near the threshold is considered to strengthen the effect of the threshold cusp. When $\gamma$ is increased, as shown by the dashed line ($\gamma=0.07$), $\sigma_{11}^{\rm N}(E)$ becomes slightly enhanced, in particular in the region below the threshold. On the other hand, when $\gamma$ is slightly decreased, $\sigma_{11}^{\rm N}(E)$ is suppressed, as seen in the dash-dotted line ($\gamma=-0.01$). In both cases, the shape of $\sigma_{11}^{\rm N}(E)$ is modified from the Flatt\'{e} amplitude only slightly, but the cusp structure is maintained. However, with a large negative $\gamma=-10.0$ (solid line), the cross section shows a qualitatively different behavior. One finds that the cusp at the threshold disappears due to the sign flip of the slope of the cross section below the threshold. Because of this, the cross section shows a small dip structure above the threshold. This dip structure is not caused by the zero point of the scattering amplitude, but it exhibits the local minimum of $\sigma_{11}^{\rm N}(E)$. It is clear that that the Flatt\'{e} amplitude cannot reproduce the cross section with the dip similar to the solid line.

\begin{figure}[tbp]
    \centering
    \includegraphics[width = 8cm, clip]{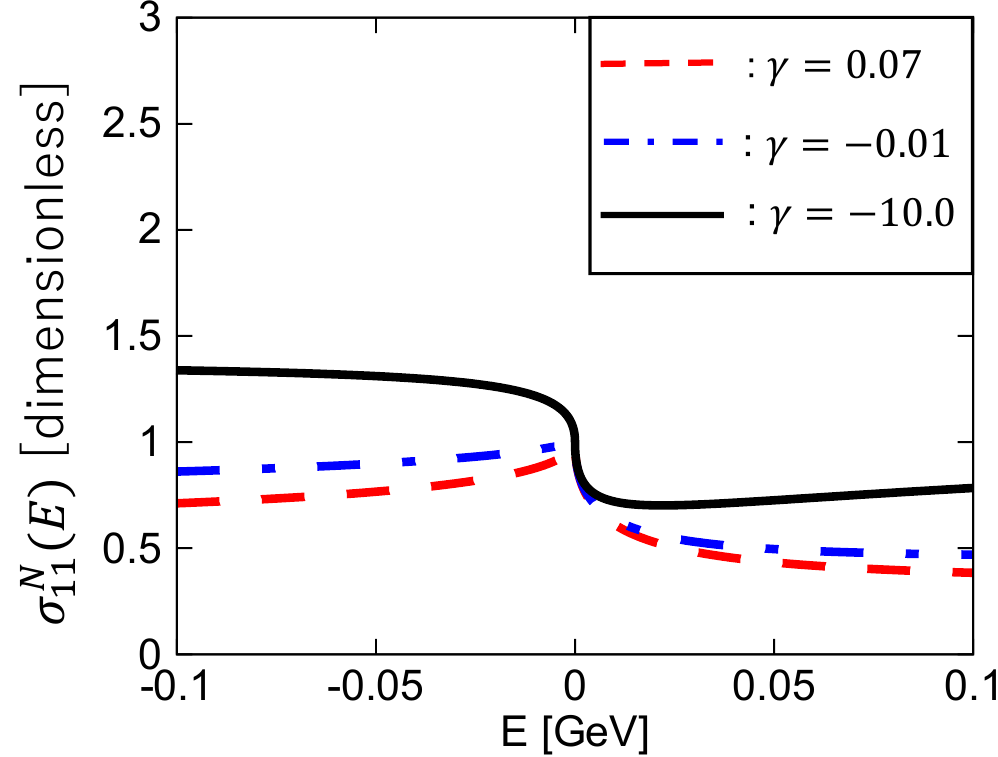}
    \caption{
    Same as Fig.~\ref{fig:sigmaG11 cusp A22+, epsilon+, aG=-1.0-i1.0} but with $A_{22}^{-}$ and $\epsilon^{-}$.
    }
    \label{fig:sigmaG11 cusp A22-, epsilon-, aG=-1.0-i1.0}
\end{figure}

The cross sections $\sigma_{11}^{\rm N}(E)$ with $A_{22}^{-}$ and $\epsilon^-$ are plotted in Fig.~\ref{fig:sigmaG11 cusp A22-, epsilon-, aG=-1.0-i1.0} for $\gamma=0.07,\ -0.01$, and $-10.0$. The cross section shows the cusp structure for $\gamma=0.07$ (dashed line) and $\gamma=-0.01$ (dash-dotted line), with a slightly enhanced strength. When $\gamma$ is taken as large and negative value ($\gamma=-10.0$), a dip appears above the threshold instead of the cusp. 

With the Flatt\'e amplitude, the cross section shows a peak below the threshold for ${\rm Re}(a_{\rm F})>0$ or a cusp at the threshold for ${\rm Re}(a_{\rm F})<0$.
From the above results, we find that the shape of the elastic scattering cross section in channel 1 can be changed significantly from these typical behaviors by varying $\gamma$. In particular, if $\gamma$ is large and negative, the peak can be changed into a cusp even with ${\rm Re}(a_{\rm G})>0$, and $\sigma_{11}^{\rm N}(E)$ can have a dip near the threshold. Therefore, in analyzing the scattering near the threshold, it is important to perform analysis paying attention to the peak, cusp, and dip structures, rather than simply fitting the data by the Flatt\'{e} amplitude~\cite{Dong:2020hxe,Baru:2024ptl}.

\section{Summary}

In this paper, we discuss the behavior of the coupled-channel scattering amplitude near the higher energy threshold by constructing the General amplitude with new parametrization. First, in Sec.~\ref{sec : coupled channel}, we introduce the Contact amplitude and Flatt\'{e} amplitude based on the effective field theory. Through the comparison of two amplitudes, we show that the origin of the problem of the number of parameters in the Flatt\'{e} amplitude near the threshold~\cite{Baru:2004xg} can be traced back to the vanishing of the determinant of the scattering amplitude. It is also shown that the standard parametrization with three parameters $a_{ij}$ in the Contact amplitude is not suitable to smoothly connect to the Flatt\'{e} amplitude. 

Based on this observation, we modify the renormalization conditions in the effective field theory and construct the General amplitude with three alternative parameters, $A_{22}, \epsilon$, and $\gamma$. With $\gamma\neq 0$, the General amplitude has one-to-one correspondence with the Contact amplitude, and it directly reduces to the Flatt\'{e} amplitude at $\gamma=0$. Namely, the General amplitude allows us to examine how the Contact amplitude approaches the Flatt\'{e} amplitude by varying the parameter $\gamma$. It is also shown in the General amplitude that the scattering length is defined in the (2,2) component (higher energy channel), which is in general different from the constant term in the denominator of the (1,1) component (lower energy channel), except for the special case of $\gamma=0$ corresponding to the Flatt\'{e} amplitude. Finally, we show that the (1,1) component of the General amplitude contains the background term in addition to the pole term, even in the linear order in the momentum. Since the background term disappears at $\gamma=0$, the parameter $\gamma$ is considered to control the existence of the background. It is shown that the background term can cause the zero of the amplitude by the interference with the pole term.

Finally, the behavior of the scattering cross section of the General amplitude is numerically studied by varying $\gamma$ with the scattering length being fixed. For a small $|\gamma|$, the cross section shows either the peak structure or the threshold cusp structure, as typically observed in the Flatt\'{e} amplitude. It is however shown that the scattering length obtained from the analysis by the Flatt\'{e} amplitude may quantitatively deviate from the exact value, due to the contribution from the background term. It is also found that by enhancing the background contribution with large and negative $\gamma$, a dip structure can be caused by the zero of the scattering amplitude (see also Refs.~\cite{Dong:2020hxe,Baru:2024ptl}), which cannot be reproduced by the simple Flatt\'{e} amplitude. In such cases, the threshold cusp structure may appear even when the real part of the scattering length is positive.

The general amplitude proposed in this study should be useful to extract the scattering length and pole position of the near-threshold exotic hadrons through the analysis of the experimental data. Since the actual exotic hadrons are often coupled to three or more channels, the extension of the General amplitude for multi-channel scattering serves as a future prospect. To enlarge the applicability of the framework, it is necessary to include the higher-order terms in the effective field theory, which also gives precise determination of the effective range. With these extensions, we expect that the application of the General amplitude to the actual experimental data will help to accurately determine the properties of the exotic hadrons near the threshold.

\begin{acknowledgments}
The authors thank Vadim Baru, Johann Haidenbauer, Yudai Ichikawa, and Kiyoshi Tanida for fruitful discussions.
 This work has been supported in part by the Grants-in-Aid for Scientific Research from JSPS (Grants
No.~JP23H05439, 
No. JP22K03637, and 
No. JP18H05402),
 by the RCNP Collaboration Research network (COREnet) 048 "Revealing the nature of exotic hadrons in Belle (II) by collaboration of experimentalists and theorists", 
and by JST SPRING, 
Grant Number JPMJSP2156

\end{acknowledgments}

\appendix
\section{$N$ channel scattering}\label{sec:Nchannel}

Here we compare the number of independent parameters in the near-threshold Contact and Flatt\'e amplitudes in the two-body scattering with $N$ channels. With a straightforward generalization of Sec.~\ref{subsec: derivation of Contact}, the $N$ channel Contact amplitude reads
\begin{align}
    [f^{\rm C}(E)]^{-1} = 
    \begin{pmatrix}
        -\frac{1}{a_{11}} - ip_1 & \frac{1}{a_{12}} & \dotsb & -\frac{1}{a_{1N}} \\
        \frac{1}{a_{12}} & -\frac{1}{a_{22}} - ip_2 & \dotsb & \frac{1}{a_{2N}} \\
        \vdots & \vdots & \ddots & \vdots \\
        -\frac{1}{a_{1N}}& \frac{1}{a_{2N}} & \dotsb & -\frac{1}{a_{NN}} - ip_N
    \end{pmatrix},
    \label{eq:the invers matrix of the EFT amplitude N}
\end{align}
where $p_j$ is the momentum in channel $j$. The amplitude $f^{\rm C}(E)$ contains $N(N+1)/2$ independent parameters $a_{ij}$. The Flatt\'e amplitude with $N$ channels can be written as 
\begin{align}
    f^{\rm F}(E) &= \frac{1}{2E_{\rm BW} - 2E - i\sum_{j=1}^Ng_j^2p_j } \notag \\
    &\quad \times
    \begin{pmatrix}
        g_1^2 & g_1g_2 & \dotsb & g_1g_N \\
        g_1g_2 & g_2^2 & \dotsb & g_2g_N \\
        \vdots & \vdots & \ddots & \vdots \\
        g_1g_N & g_2g_N & \dotsb & g_N^2
    \end{pmatrix}, \label{eq: the Flatte derived from EFT of N channel}
\end{align}
which contains $N$ coupling constants $g_j(j=1,\dotsb N)$ and one bare energy $E_{\rm BW}$, so there are $N+1$ independent parameters in total. Near the threshold of channel 1, by neglecting the terms with $p_1^2$ or higher, we obtain
\begin{align}
    f^{\rm F}(E) &\sim  \frac{1}{\alpha  - ip_1 - i\sum_j r_j p_j^{(0)}}
    \nonumber \\
    &\quad \times 
    \begin{pmatrix}
        1 & \sqrt{r_2} & \dotsb & \sqrt{r_N} \\
        \sqrt{r_2} & r_2 & \dotsb & \sqrt{r_2r_N} \\
        \vdots & \vdots & \ddots & \vdots \\
        \sqrt{r_N} & \sqrt{r_2r_N} & \dotsb & r_N
    \end{pmatrix},
    \label{eq:the Flatte amplitude up to first order in k which is represented by R and alpha N}
\end{align}
where we define 
\begin{align}
    \frac{g_j^2}{g_1^2} &= R_j , \quad
    \frac{2E_{\rm BW}}{g_1^2} = \alpha, \\
    p^{(0)}_j & =p_j(p_1=0), 
\end{align}
In this case, the amplitude has $N$ parameters, $r_j(j=2,\dotsb , N)$ and $\alpha$. 

As summarized in Table~\ref{tbl:parameters}, the Contact amplitude always has a larger number of parameters. Namely, there should be 
\begin{align}
    \frac{1}{2}N(N-1),
\end{align}
conditions imposed on the Flatt\'e amplitude, as in the case of two-channel scattering. This is achieved by demanding all the cofactors of the off-diagonal components vanish.

\begin{table}
\caption{Number of independent parameters in the Contact amplitude~\eqref{eq:the invers matrix of the EFT amplitude N} and the Flatt\'e amplitude~\eqref{eq:the Flatte amplitude up to first order in k which is represented by R and alpha N}. \label{tbl:parameters}}
\begin{ruledtabular}
\begin{tabular}{llll}
Channel     & 2 & 3 & $N$ \\
\hline
Contact~\eqref{eq:the invers matrix of the EFT amplitude N}     & 3 & 6 & $N(N+1)/2$ \\
Flatt\'e~\eqref{eq:the Flatte amplitude up to first order in k which is represented by R and alpha N}    & 2 & 3 & $N$ \\
Conditions  & 1 & 3 & $N(N-1)/2$ \\
\end{tabular}
\end{ruledtabular}
\end{table}

\bibliography{refs.bib}

\end{document}